\documentclass[12pt]{article}
\usepackage{amssymb,amsmath,epsfig}
\allowdisplaybreaks
\begin{document}
\title{\bf Anisotropic Compact Stellar Objects in Modified Gauss-Bonnet Gravity}
\author{M. Sharif \thanks{msharif.math@pu.edu.pk} and Amna Ramzan
\thanks{amnaramzan2168@gmail.com}\\
Department of Mathematics, University of the Punjab,\\ Quaid-e-Azam
Campus, Lahore-54590, Pakistan.}
\date{}

\maketitle

\begin{abstract}
This paper is devoted to studying anisotropic compact stellar
structures by adopting embedding class-1 technique in the background
of modified Gauss-Bonnet gravity. The unknown constants are
evaluated by the matching of interior spacetime with the
Schwarzschild exterior geometry corresponding to
$f(\mathcal{G})=\chi\mathcal{G}^{n}$ model, where $\chi$ and $n$ are
positive constants. The observed masses of compact star candidates
(SAX J1808.4-3658, Vela X-1, PSR J0348+0432, 4U 1608-52) are used
with the condition of vanishing radial pressure at the stellar
surface to predict their radii. We have examined viability and
stability of the resulting solution through graphical behavior of
matter variables, energy constraints, adiabatic index and causality
condition. It is found that embedding class-1 solution for
anisotropic compact stars is viable and stable in this theory.
\end{abstract}

\section{Introduction}

In the present age, the cosmological and astrophysical scenarios
have motivated many researchers to discuss the universe and its
mysterious constituents. Stars are considered as the fundamental
component of galaxy as well as the main ingredient of astrophysics.
The fusion reactions play vital role in the structure formation and
evolution of astronomical objects. The fusion process generates the
outward directed pressure inside a star which is balanced by the
inward directed force of gravity to keep the star in equilibrium
state. Once the nuclear fuel is burnt out completely, there appears
no enough pressure to stop the star collapse. This leads to the
formation of new compact objects categorized as white dwarfs,
neutron stars and black holes.

The interior geometry of compact stars provokes the researchers to
study the surprising characteristics of celestial objects.  The
presence of anisotropy in spherically symmetric objects influences
important physical characteristics of relativistic objects. Ruderman
\cite{1'} proposed that nuclear matter possesses anisotropy if the
matter density for relativistic object is equal to $10^{15}g/cm^3$.
The pressure anisotropy in matter distribution occurs due to
viscosity, phase transition \cite{1a}, pion condensation \cite{1b}
and super fluid \cite{1c}. Many people studied the effects of
anisotropy on mass, radius and redshift of the stars by considering
the radial and tangential components of pressure.  Herrera and
Santos \cite{1} analyzed the causes and effects of anisotropy for
self-gravitating system. Harko and Mak \cite{2} studied the interior
solutions for anisotropic celestial objects and observed their
physical features. Dev and Gleiser \cite{3} examined the stability
of anisotropic stellar objects in Newtonian and general relativistic
unit. Hossein et al. \cite{4} analyzed the stability of anisotropic
objects along with the effects of cosmological constant by adopting
Krori-Barua solution. Paul and Deb \cite{5} found feasible solutions
for compact objects under the influence of pressure anisotropy.

Different techniques, such as constraints on the matter constituents
or a particular form of equation of state, are used to formulate the
interior solutions of stellar models. The embedding of curved
geometry into higher-dimensional spacetime also provides a relation
between two potentials (temporal and radial) which helps in finding
a solution of the field equations. The technique in which
$n$-dimensional manifold $\mathcal{M}_{n}$ is embedded into
$(n+k)$-dimensional manifold $\mathcal{M}_{k}$ is named as embedding
class-$k$ for $\mathcal{M}_{n}$, where $k$ is the minimum number of
extra dimension. Eisenhart \cite{6} defined the embedding class-1
condition to find solutions for static spherically symmetric
geometry. Maurya et al. \cite{7} found a new star model of embedding
class-1 with different potential functions. The same authors
\cite{8} discussed the stability and physical characteristics of
anisotropic compact objects by using embedding class-1 technique.
Singh and Pant \cite{9} used the same approach to find solutions
that describe the internal geometry of astrophysical objects. Bhar
et al. \cite{10} found anisotropic solutions through this technique
and examined the behavior of different compact stellar models. Singh
et al. \cite{11} analyzed a new model of compact stars (free from
geometric singularity) by the embedding of 4-dimensional space into
5-dimensional pseudo-Euclidean geometry.

The mysteries behind the current accelerated expansion of cosmos has
attracted the attention of modified gravitational theories. These
theories have widely been employed to provide the information about
hidden aspects of dark energy and dark matter. The $f(\mathcal{R})$
theory is considered as the simplest extension to general
relativity, obtained by using generic function $f(\mathcal{R})$ of
Ricci scalar ($\mathcal{R}$) in the Einstein-Hilbert action.
Capozziello et al. \cite{13a} investigated the hydrostatic
equilibrium of stellar structures in the framework of
$f(\mathcal{R})$ gravity. Nojiri and Odintsov \cite{13x} inspected
the compatibility of modified theories with local tests. Olmo
\cite{13y} applied Palatini approach in the context of
$f(\mathcal{R})$ and $f(\mathcal{R,Q})$ theories to analyze the
aspects related to dark energy, dark matter and quantum gravity.
Zubair and Abbas \cite{13b} discussed the mathematical modeling of
compact stars with static anisotropic interiors by considering the
Krori-Barua spacetime in $(\mathcal{R},T)$ gravity. Yousaf et al.
\cite{13c} explored the role of different forces as well as equation
of state parameter on a viable configuration of anisotropic
spherical structures in the same theory.

Nojiri and Odintsov \cite{12} introduced another modified theory
referred to as $f(\mathcal{G})$ gravity, where $\mathcal{G}$
represents Gauss-Bonnet invariant term given by
$\mathcal{G}=\mathcal{R}^2-4R_{\alpha\beta}\mathcal{R}^{\alpha\beta}+\mathcal{R}_{\alpha\beta\xi\eta}
\mathcal{R}^{\alpha\beta\xi\eta}$. They discussed different
cosmological aspects that describe the accelerated expansion of the
universe and possible phase transition from deceleration to
acceleration in $f(\mathcal{R})$, $f(\mathcal{G})$ and
$f(\mathcal{R},\mathcal{G})$ theories \cite{13d}. Bamba et al.
\cite{13} analyzed the finite future time singularities in
$f(\mathcal{R},\mathcal{G})$ as well as $f(\mathcal{G})$ theory and
studied possible solutions by involving higher-order curvature
terms. Nojiri and Odintsov \cite{13e} also studied various relations
between modified theories and concluded that such theories may
provide the description of inflation with dark energy epoch.
Oikonomou \cite{13e'} discussed bounce cosmology with a Type IV
singularity at the bouncing point in the framework of
$f(\mathcal{G})$ gravity. He examined that mimetic vacuum
$f(\mathcal{G})$ gravity can explain the singular bounce cosmology.
Nojiri et al. \cite{13e''} discussed some issues and the
developments of modified gravity such as late time acceleration,
inflation and bouncing cosmology. They demonstrated that
$f(\mathcal{R})$, $f(\mathcal{G})$ and $f(T)$ theories provide the
required description of the universe.

Felice and Tanaka \cite{13'} analyzed the behavior of perturbations
in the background of $f(\mathcal{R,G})$ gravity and showed that the
existence of ghost is inevitable for generic models. Moreover, they
discussed some special cases which avoid the existence of ghost.
Nojiri et al. \cite{13''} proposed a technique for Gauss-Bonnet
theories ($f(\mathcal{G})$ and $f(\mathcal{R,G})$) to eliminate the
ghost modes in the equations of motion. The variation of action
integral with respect to perturbed metric tensor yields the
perturbed field equations. It is shown that these equations contain
fourth order derivatives of the metric with respect to time
coordinate. Therefore, ghost modes might appear in $f(\mathcal{G})$
model. They applied constraints on the action integral and used the
technique of Lagrange multiplier to produce ghost-free modes. Nojiri
et al. \cite{13'''} studied the inflationary aspects of ghost-free
$f(\mathcal{G})$ gravity. By using the method of slow-roll
approximation, they calculated the slow-roll and observational
indices. The results were consistent with the latest Planck data
\cite{13''}. Furthermore, they discussed cosmological evolution in
the presence of exponentially evolving Hubble rate by considering a
coupling function. They demonstrated that the ghost-free model can
produce inflationary phenomenon which is compatible with the
observational data \cite{13'''}.

Abbas et al. \cite{14} inspected the anisotropic behavior of compact
stars using Krori-Barua spacetime in $f\mathcal{(G)}$ gravity and
analyzed physical characteristics for different star models. Sharif
and Fatima \cite{15} investigated the isotropic and anisotropic
spherical symmetric solutions in this theory by assuming linear
equation of state (EoS). Sharif and Naz \cite{16} examined the
gravitational collapse in the same scenario with electromagnetic
effects and concluded that there is a decrease in collapsing rate
due to modified terms. Sharif and Saba \cite{17} studied the
anisotropic solutions of compact stellar objects with Karmarkar
constraint in the same theory. The same authors \cite{18} also
discussed the effects of charge on gravitational decoupled sources
and checked the stability as well as viability of the obtained
solutions.

Odintsov \cite{18a} proposed a new technique to discuss the
equations of motion, slow-roll and the corresponding observational
indices in Einstein-Gauss-Bonnet theories of gravity. Oikonomou and
Fronimos \cite{18b} discussed the non-minimal kinetic coupling
corrected Einstein-Gauss-Bonnet theories by considering the GW17017
event with the assumption that the gravitational wave propagates at
the speed of light. They also showed that GW170817 is compatible
with the recent Planck data \cite{13"}. Astashenok \cite{18c}
studied the data compiled by the LIGO collaboration \cite{18d} for
the GW190814 event and showed that this event can be represented
through some models in the background of extended theories of
gravity. They consider some particular $f(\mathcal{R})$ models to
study rotating stars and found that the obtained results were
consistent with the findings LIGO. Recently, Shamir and Naz
\cite{19} inspected some compact stellar objects corresponding to
Tolman-Kuchowicz spacetime in the background of $f(\mathcal{G})$
theory. They consider the observed data of Cen X -3, EXO 1785-248
and LMC X - 4 star models.

In this paper, we explore anisotropic spherical solution by using
embedding class-1 technique for a specific $f(\mathcal{G})$ model.
We discuss features like evolution of matter variables, mass-radius
relation, the Tolman-Oppenheimer-Volkoff (TOV) equation, energy
bounds, compactness factor, redshift parameter and stability. The
paper has the following format. Section \textbf{2} formulates the
field equations and solutions using embedding class-$1$ condition.
In section \textbf{3}, we evaluate unknown constants by smooth
matching of interior and exterior geometries at the boundary.
Section \textbf{4} explores physical features of the obtained
solution for some specific compact star models. In the last section,
we summarize our results.

\section{Field Equations for Anisotropic Source}

The action for $f(\mathcal{G})$ theory is defined as \cite{12}
\begin{equation}\label{1}
\mathcal{A}_{f(\mathcal{G})}=\int\Big(\frac{\mathcal{R}+f(\mathcal{G})}
{2\kappa^{2}}+\mathcal{L}_{m}\Big)\sqrt{-g}{d^4}x,
\end{equation}
where $f(\mathcal{G})$ is the arbitrary function of Gauss-Bonnet
invariant, $g$ represents determinant of the metric tensor
($g_{\alpha\beta}$) and $\mathcal{L}_{m}$ is the Lagrangian density
of matter. Variation of this action with respect to
$g_{\alpha\beta}$ leads to
\begin{eqnarray}\nonumber
G_{\alpha\beta}&=&\kappa^2T_{\alpha\beta}+\frac{1}{2}g_{\alpha\beta}f(\mathcal{G})
-(2\mathcal{R}\mathcal{R}_{\alpha\beta}-4\mathcal{R}_{\alpha}^{\xi}\mathcal{R}_{\xi\beta}-4\mathcal{R}_{\alpha\xi
\beta\eta}\mathcal{R}^{\xi\eta}+2\mathcal{R}_{\alpha}^{\xi\eta\delta}\\\nonumber&\times&
\mathcal{R}_{\beta\xi\eta\delta})f_\mathcal{G}(\mathcal{G})-(2\mathcal{R}g_{\alpha\beta}\nabla^2
-2\mathcal{R}\nabla_{\alpha}\nabla_{\beta}-4g_{\alpha\beta}\mathcal{R}^{\xi\eta}
\nabla_\xi\nabla\eta-4\mathcal{R}_{\alpha\beta}\nabla^2\\\label{2}&+&4
\mathcal{R}^\xi_\alpha\nabla_{\beta}\nabla_{\xi}+4\mathcal{R}^{\xi}_{\beta}
\nabla_{\alpha}\nabla_{\xi}+4\mathcal{R}_{\alpha\xi\beta\eta}
\nabla^\xi\nabla^\eta)f_\mathcal{G}(\mathcal{G}),
\end{eqnarray}
where $f_\mathcal{G}({\mathcal{G}})$ is the derivative with respect
to $\mathcal{G}$,
$\nabla^{2}=g^{\alpha\beta}\nabla_{\alpha}\nabla_{\beta}$  and
$\nabla_{\alpha}$ denotes the covariant derivative. The
stress-energy tensor is expressed by
\begin{equation}\label{3}
T_{\alpha\beta}=g_{\alpha\beta}\mathcal{L}_{m}
-2\frac{\partial\mathcal{L}_{m}}{\partial g^{\alpha\beta}}.
\end{equation}
We take anisotropic matter distribution of stellar objects described
by
\begin{equation}\label{4}
T_{\alpha\beta}={(\rho+p_{t})\mathcal{V}_{\alpha}\mathcal{V}_{\beta}}
-p_{t}g_{\alpha\beta}+{(p_{r}-p_{t})\mathcal{U}_{\alpha}\mathcal{U}_{\beta}},
\end{equation}
where $\rho$, $ p_{t}$ and $p_{r}$ represent the energy density,
tangential and radial pressures, respectively. Also,
$\mathcal{V}_{\alpha}$ indicates the four-velocity and
$\mathcal{U}_{\alpha}$ represents the four-vector in radial
direction in comoving frame that satisfy
\begin{equation*}
\mathcal{V}^{\alpha}\mathcal{V}_{\alpha}=1,\quad
\mathcal{U}_{\alpha}\mathcal{U}^{\alpha}=-1.
\end{equation*}

The most interesting feature of $f(\mathcal{G})$ theory is that it
may neglect the ghost terms and regularize the action due to
Gauss-Bonnet invariant. In order to study the structural properties
of compact objects, we take a power-law model of $f(\mathcal{G})$
theory \cite{20}
\begin{equation}\label{5}
f(\mathcal{G})=\chi \mathcal{G}^{n},
\end{equation}
where $\chi$ denotes the arbitrary constant and $n>0$ with $n\neq1$.
Here, we choose $\chi=1$ and $n=2$. The action (\ref{1}) with the
considered form of $f(\mathcal{G})$ is compatible with the
observational data of the expanding universe \cite{6b}. The generic
function $f(\mathcal{G})$ is viable, since it is compliant with
solar system constraints \cite{20}. Moreover, any $f(\mathcal{G})$
model is said to be viable if the generic function and its
derivatives are regular. These conditions are fulfilled by the
considered model. To discuss the internal geometry of compact stars,
we consider static and spherically symmetric metric
\begin{equation}\label{6}
ds^{2}=e^{\nu({r})}dt^{2}-e^{\lambda({r})}dr^{2}-r^{2}{(d\theta^{2}
+\sin^{2}\theta d\phi^{2})},
\end{equation}
where $\lambda(r)$ and $\nu(r)$ represent the gravitational metric
potentials which depend on radial coordinate $r$ only. Equations
(\ref{2}) and (\ref{4})-(\ref{6}) yield the following field
equations
\begin{eqnarray}\nonumber
e^{-\lambda }\big(\frac{\lambda '}{r}-\frac{1}{r^2}\big)
+\frac{1}{r^2}&=&\frac{\mathcal{G}^2}{2}+8\pi\rho+\Big[\frac{e ^{-2
\lambda }}{r^{2}}\Big(2(e^{\lambda }-3)\\\nonumber&\times&(2
\mathcal{G}^{' } \lambda '-\mathcal{G} \lambda ' \nu ')+2(e^{\lambda
}-1)(2\\\label{7}&\times&\mathcal{G} \nu ''+\mathcal{G}(\nu ')^2-4
\mathcal{G}^{''})\Big)\Big],
\\\nonumber
e^{-\lambda }\big(\frac{1}{r^2}+\frac{\nu '}{r}\big)
-\frac{1}{r^2}&=&-\frac{\mathcal{G}^2}{2}+8\pi p_r+\Big[\frac{e ^{-2
\lambda}}{r^{2}}\Big(2(e^{\lambda }\\\nonumber&-&3)(\mathcal{G}
\lambda '\nu'+2 \mathcal{G}^{'} \nu ') -2 (e^{\lambda
}-1)\\\label{8}&\times&(2 \mathcal{G} \nu ''+\mathcal{G} (\nu
')^2)\Big)\Big],
\\\nonumber
\frac{e^{-\lambda }}{2}\big(-\frac{\lambda'\nu'}{2}+\nu''
+\frac{\nu'^2}{2}+\frac{\nu'-\lambda '}{r}\big)&=&
-\frac{\mathcal{G}^2}{2}+8\pi p_t+\Big[\frac{e^{-2 \lambda}}{r^{2}}
\Big(2\mathcal{G}\\\nonumber&\times&(e^{\lambda }-3) \lambda'\nu
'-2(e^{\lambda }-1)(2\mathcal{G}\nu''\\\nonumber&+& \mathcal{G}
(\nu')^2)-2 r\nu'(2\mathcal{G}^{''}-3 \mathcal{G}^{'} \lambda
')\\\label{9}&-&2 \mathcal{G}^{'}r(2\nu ''+(\nu ')^2)\Big)\Big],
\end{eqnarray}
where $(')$ manifests differentiation with respect to the radial
coordinate. Solving these equations, we have
\begin{eqnarray}\nonumber
\rho&=&-\frac{1}{16\pi
r^2}e^{-2\lambda}\Big(\mathcal{G}^2e^{2\lambda}r^2-4\mathcal{G}
e^{\lambda}\lambda'\nu'+12\mathcal{G}\lambda'\nu'+8\mathcal{G}^{'}
e^{\lambda}\lambda'\\\nonumber&-&24\mathcal{G}^{'}\lambda'+8\mathcal{G}
e^{\lambda}\nu''+4\mathcal{G}e^{\lambda}(\nu')^2-16\mathcal{G}^{''}
e^{\lambda}-8\mathcal{G}\nu''-4\mathcal{G}(\nu')^2\\\label{10}&+&16
\mathcal{G}^{''}+2e^{\lambda}-2e^{2\lambda}-2e^{\lambda}r\lambda'\Big),
\\\nonumber
p_{r}&=&\frac{1}{16 \pi  r^2}e^{-2 \lambda }\Big(\mathcal{G}^2 e^{2
\lambda } r^2-4 \mathcal{G} e^{\lambda } \lambda ' \nu '+12
\mathcal{G} \lambda ' \nu '+8 \mathcal{G} e^{\lambda } \nu ''+4
\mathcal{G} e^{\lambda } (\nu ')^2\\\label{11}&-&8\mathcal{G}^{'}
e^{\lambda } \nu '-8 \mathcal{G} \nu ''-4 \mathcal{G} (\nu ')^2+24
\mathcal{G}^{'} \nu '+2 e^{\lambda }-2 e^{2 \lambda }+2 e^{\lambda }
r \nu '\Big),
\\\nonumber
p_{t}&=&\frac{1}{{32 \pi r^2}}e^{-2 \lambda } \Big(2 \mathcal{G}^2
e^{2 \lambda } r^2-8 \mathcal{G} e^{\lambda } \lambda ' \nu
'+24\mathcal{G} \lambda ' \nu '+16 \mathcal{G} e^{\lambda } \nu
''-{16 \mathcal{G} \nu ''}\\\nonumber&+&8 \mathcal{G} e^{\lambda }
(\nu ')^2-8\mathcal{G} (\nu ')^2-24\mathcal{G}^{'} r \lambda ' \nu
'+16 \mathcal{G}^{'} r \nu ''+8\mathcal{G}^{'} r (\nu ')^2+16
\mathcal{G}^{''} r \nu '\\\label{12}&-&e^{\lambda } r^2 \lambda '
\nu '+2 e^{\lambda } r^2 \nu ^{''}+e^{\lambda } r^2 (\nu ')^2-2
e^{\lambda } r \lambda '+2 e^{\lambda } r \nu ' \Big).
\end{eqnarray}
The mass of a spherical object is one of the salient features which
determines the total energy within the sphere. The definition of
energy content within a given piece of the fluid distribution is not
unique due to ambiguity in the localization of energy \cite{14''}.
This problem exists in modified theories as well. However,
researchers have used the definitions proposed by Tolman
\cite{14'''} as well as Misner-Sharp \cite{21} to describe the mass
of the distribution in modified theories. Hence, in the present
work, mass function is characterized through Misner-Sharp definition
as
\begin{equation*}
m{(r)}=\frac{r}{2}(1+{g^{\delta\sigma}}r_{,\delta}r_{,\sigma}),
\end{equation*}
which gives
\begin{equation}\label{13}
e^{-\lambda(r)}=1-\frac{2m}{r}.
\end{equation}

An $n$-dimensional spherical line element always belongs to
embedding class-$2$. However, it can be embedded in an
$n+1$-dimensional space if a symmetric tensor $b_{\xi\eta}$
satisfies the Gauss-Codazi equations
\begin{equation}\label{14}
\mathcal{R}_{\xi\eta\mu\nu}=2eb_{\xi{[}\mu}b_{\nu{]}\eta}\quad
\textrm{and} \quad
b_{\xi{[}\eta;\mu{]}}-\Gamma^{\lambda}_{\eta\mu}b_{\xi\lambda}
+\Gamma^{\lambda}_{\xi{[}\eta}b_{\mu{]}\lambda}=0.
\end{equation}
 Here, $e=\pm1$ and $b_{\xi\eta}$ are the coefficients of second
differential form. Eiesland \cite{22} evaluated a necessary and
sufficient condition for embedding class-$1$ as
\begin{equation}\label{15}
\mathcal{R}_{0101}\mathcal{R}_{2323}=\mathcal{R}_{0202}\mathcal{R}_{1313}-\mathcal{R}_{1202}\mathcal{R}_{1303},
\end{equation}
and hence
\begin{equation}\label{16}
{(\lambda'-\nu')\nu^{'}e^{\lambda}}+2{(1-e^{\lambda})}\nu^{''}+{\nu^{'2}}=0.
\end{equation}
Since both the metric potentials are unknown, so we assume one of
them such that \cite{8}
\begin{equation}\label{17}
\nu{(r)}=2\alpha r^{2}+\ln{\gamma},
\end{equation}
where $\alpha$ and $\gamma$ are positive constants. The existence of
geometric as well as physical singularities inside the stellar
models is a significant aspect in the analysis of compact
structures. For this purpose, we investigate the behavior of metric
function at the center of star. It is well-known that for physical
validity of the solution, the metric potential must be positive,
regular and monotonically increasing function within the entire
stellar model \cite{23}. We note that
\begin{equation*}
\nu'(r)=4\alpha r  \quad \textrm{and} \quad \nu''(r)=4\alpha.
\end{equation*}
It follows that $\nu(0)=\gamma$, $\nu'(0)=0$ and $\nu^{''}(0)>0$ at
the center. Hence, the chosen metric potential is free from
singularity and monotonically increasing function having minimum at
the center of the star.
 We use embedding
class-1 approach to obtain $\lambda(r)$ from Eq.$(\ref{18})$ as
\begin{equation}\label{18}
\lambda{(r)}=\ln{(1+{\beta\nu^{'2}}e^{\nu})},
\end{equation}
where $\beta$ is the integration constant. The behavior of metric
functions at the center of compact objects is presented in Figure
\textbf{1} which shows that the metric potentials satisfy the
required physical conditions. We note that both the metric functions
are minimum at the center and attain large values at the boundary of
stars.
\begin{figure}\center
\epsfig{file=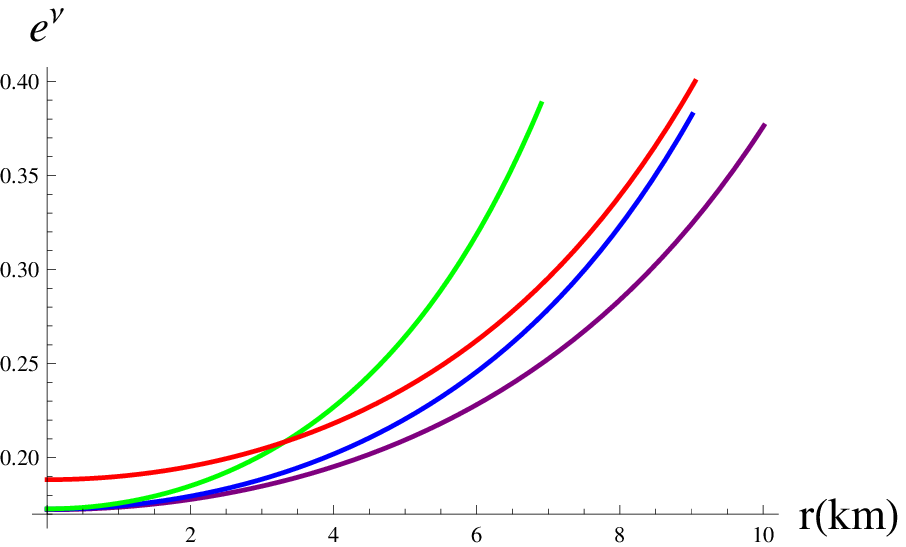,width=0.43\linewidth}
\epsfig{file=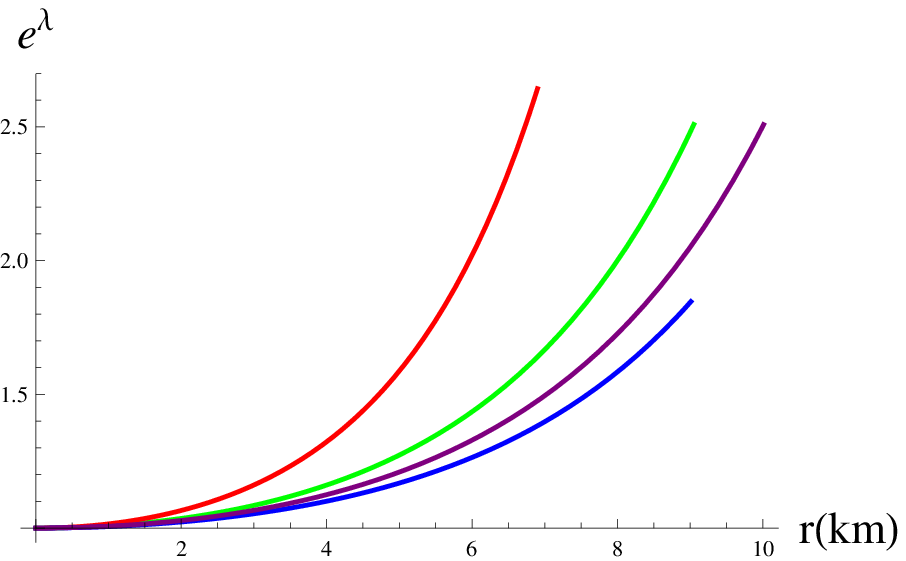,width=0.43\linewidth} \caption{Plots of metric
functions for  Vela X-1 (green), SAX J1808.4-3658 (red), 4U 1608-52
(blue) and PSR J0348+0432 (purple) versus radial coordinate.}
\end{figure}

\section{Matching with Exterior Metric}

The set of constants $(\alpha,\beta,\gamma)$ interprets physical
features as well as geometry of anisotropic stellar objects that can
be evaluated by the smooth matching of interior ($\mathcal{M^-}$)
and exterior ($\mathcal{M^+}$) spacetimes  on the boundary $(\sum)$
of the star. In general relativity, Jebsen-Birkhoff's theorem states
that the gravitational field outside a spherically symmetric object
is static. To define the exterior region of the compact stars,
Schwarzschild solution is considered as the most appropriate choice.
However, an exterior vacuum solution has not been evaluated in the
context of $f(\mathcal{G})$ theory so many researchers
\cite{14,15,17} have used Schwarzschild metric to describe the
exterior region of compact stars. Therefore in the present work, the
external region is defined by the Schwarzschild metric as
\begin{equation}\label{19}
ds^{2}={(1-\frac{2M}{r})}dt^{2}-{(1-\frac{2M}{r})}^{-1}dr^{2}-r^{2}
{(d\theta^{2}+\sin^{2}\theta d\phi^{2})},
\end{equation}
where $M$ is the total mass. The continuity of metric coefficients
at the boundary gives
\begin{equation}\label{20}
e^{\nu{(R)}}=\gamma e^{2\alpha R^{2}}=1-\frac{2M}{R}, \quad
e^{\lambda(R)}=1+16\alpha^{2}\beta\gamma R^{2}e^{2\alpha
R^{2}}=(1-\frac{2M}{R})^{-1}.
\end{equation}
Using Eqs.(\ref{17}) and (\ref{20}), we obtain
\begin{equation}\label{21}
\alpha=\frac{M}{2R^{2}(R-2M)},\quad \beta=\frac{R^{3}}{2M},\quad
\gamma=\frac{(R-2M)}{R}e^{\frac{M}{(2M-R)}}.
\end{equation}
Substituting Eqs.(\ref{17}) and (\ref{18}) into
(\ref{10})-(\ref{12}), the energy density, radial and tangential
pressures are
\begin{eqnarray}\nonumber
\rho&=&\frac{1}{16 \pi  r (16 \alpha^2 \beta \gamma r^2 e^{2 \alpha
r^2}+1)^3}\Big[r (-1536 \alpha^3 \beta \mathcal{G}\gamma e^{2 \alpha
r^2} (2 \alpha r^2+1)-\mathcal{G}^2\\\nonumber&\times&(16 \alpha^2
\beta \gamma r^2 e^{2 \alpha r^2}+1)^3+256 \alpha^2 \beta
\mathcal{G}^{''} \gamma e^{2 \alpha r^2}(16 \alpha^2 \beta\gamma r^2
e^{2 \alpha r^2}+1)\\\nonumber&+&32 \alpha^2 \beta \gamma e^{2\alpha
r^2} (256 \alpha^4 \beta^2 \gamma^2 r^4 e^{4\alpha
r^2}+64\alpha^3\beta \gamma r^4 e^{2 \alpha r^2}+64 \alpha^2\beta
\gamma r^2 e^{2\alpha r^2}\\\label{22}&+&4\alpha r^2+3))-512
\alpha^2 \beta \mathcal{G}^{'} \gamma e^{2 \alpha r^2}(2 \alpha
r^2+1)(8 \alpha^2 \beta \gamma r^2 e^{2 \alpha r^2}-1)\Big],
\\\nonumber
p_{r}&=&\frac{1}{{16 \pi  r (16 \alpha^2 \beta \gamma r^2 e^{2
\alpha r^2}+1)^3}}\Big[r (4096\alpha^6 \beta^3 \gamma^3 r^4 e^{6
\alpha r^2} (\mathcal{G}^2 r^2-2)\\\nonumber&+&2048\alpha^5 \beta^2
\gamma^2 r^4 e^{4 \alpha r^2}+256\alpha^4 \beta \gamma r^2 e^{2
\alpha r^2}(3 \beta \mathcal{G}^2 \gamma r^2 e^{2\alpha r^2}-4 \beta
\gamma e^{2 \alpha r^2}\\\nonumber&+&12 \mathcal{G})+256
\alpha^3\beta\gamma e^{2 \alpha r^2} (6 \mathcal{G}+r^2)+16 \alpha^2
\beta \gamma e^{2 \alpha r^2} (3\mathcal{G}^2 r^2-2)+8
\alpha\\\label{23}&+&\mathcal{G}^2)-64 \alpha\mathcal{G}^{'}(128
\alpha^4 \beta^2 \gamma^2 r^4 e^{4\alpha r^2}-8 \alpha^2 \beta
\gamma r^2 e^{2 \alpha r^2}-1)\Big],
\\\nonumber
p_{t}&=&\frac{1}{16 \pi r (16 \alpha^2 \beta r^2 e^{2 \alpha
r^2}+1)^3}\Big[r (4096 \alpha^6 \beta^3 \mathcal{G}^2 \gamma^3 r^6
e^{6\alpha r^2}+128 \alpha^4 \beta \gamma r^2\\\nonumber&\times&e^{2
\alpha r^2}(6 \beta \mathcal{G}^2 r^2 e^{2 \alpha r^2}-4 \beta\gamma
e^{2 \alpha r^2}+24\mathcal{G}+r^2)+128 \alpha^3 \beta e^{2 \alpha
r^2} (12 \mathcal{G}\\\nonumber&+&r^2)+8 \alpha^2 (2 \beta \gamma
e^{2 \alpha r^2} (3 \mathcal{G}^2 r^2-2)+r^2)+32 \alpha
\mathcal{G}^{''} (16\alpha2 \beta r^2 e^{2 \alpha
r^2}\\\label{24}&+&1)+8 \alpha+\mathcal{G}^2)-32 \alpha
\mathcal{G}^{'} (2\alpha r^2+1) (32 \alpha^2 \beta \gamma r^2 e^{2
\alpha r^2}-1)\Big],
\end{eqnarray}
where
\begin{eqnarray}\nonumber
\mathcal{G}&=&-\frac{768 \alpha^3 \beta \gamma e^{ 2\alpha r^2} (1 +
2 \alpha r^2)}{(1 + 16 \alpha^2 \beta \gamma e^{2 \alpha r^2}
r^2)^3},
\\\nonumber
\mathcal{G}^{'}&=&\frac{1}{(1 + 16 \alpha^2\beta \gamma e^{2 \alpha
r^2} r^2)^4}\Big[6144 \alpha^4 \beta \gamma e^{2 \alpha r^2} r (-1 +
12 \alpha \beta \gamma e^{2 \alpha r^2} \\\nonumber&-&\alpha r^2 +
32 \alpha^2 \beta \gamma e^{2\alpha r^2} r^2 + 32 \alpha^3 \beta
\gamma e^{2 \alpha r^2} r^4)\Big],
\\\nonumber
\mathcal{G}^{''}&=&\frac{-1}{(1 + 16 \alpha^2 \beta \gamma e^{2
\alpha r^2} r^2)^5}\Big[ 6144\alpha^4 \beta \gamma e^{ 2 \alpha r^2}
(1 + 5632 \alpha^5 \beta^2\gamma^2 e^{4 \alpha r^2} r^6
\\\nonumber&+& 4096 \alpha^6 \beta^2 \gamma^2 e^{4 \alpha r^2} r^8 +
16 \alpha^3 \beta \gamma e^{2 \alpha r^2} r^2 (84 \beta \gamma e^{2
\alpha r^2} - 43 r^2) + 64 \alpha^4\\\nonumber&\times& \beta \gamma
e^{2 \alpha r^2} r^4 (64 \beta \gamma e^{2 \alpha r^2} - 7 r^2) + 4
\alpha^2 r^2 (-76 \beta \gamma e^{2 \alpha r^2} + r^2) +
\alpha\\\nonumber&\times& (-12 \beta \gamma e^{2 \alpha r^2} + 7
r^2))\Big].
\end{eqnarray}
By using the Eqs.(\ref{21}) and (\ref{23}) with the condition
$p_{r}(R)=0$, we obtain the total mass of the compact stars as
\begin{equation*}
M=\frac{1}{22}\Big[13R-\frac{7R^{2}\sqrt[3]{3}}{(-559R^{3}+44\sqrt{163}
R^{3})^{\frac{1}{3}}}+\frac{(-559R^{3}+44\sqrt{163}R^{3})^\frac{1}{3}}{\sqrt[3]{3}}\Big],
\end{equation*}

In anisotropic system, matter variables, i.e., matter density and
pressure components are often related through EoS given as
\begin{equation*}
\omega_r=\frac{p_r}{\rho}, \quad \omega_t=\frac{p_t}{\rho},
\end{equation*}
where $\omega_r$ and $\omega_t$ represent the EoS parameters. The
presence of radial and tangential pressures in the stellar object
yields the anisotropy which can be calculated from Eqs.(\ref{23})
and (\ref{24}) as
\begin{eqnarray}\nonumber
\Delta&=& p_{t}-p_{r}=\frac{1}{2 \pi r (16 \alpha^2 \beta \gamma r^2
e^{2 \alpha r^2}+1)^3}\Big[\alpha (r (16 \alpha^2 \beta \gamma r^2
e^{2\alpha r^2}+1)(\alpha r^2\\\nonumber&\times&(1-8 \alpha \beta
\gamma e^{2 \alpha r^2})^2+4 \mathcal{G}^{''})+4 \mathcal{G}^{'}
(256 \alpha^4 \beta^2 \gamma^2 r^4 e^{4 \alpha r^2}-64
\alpha^3\\\label{26}&\times&\beta \gamma r^4 e^{2\alpha r^2}-48
\alpha^2 \beta \gamma r^2 e^{2\alpha r^2}+2 \alpha r^2-1))\Big].
\end{eqnarray}

\section{Physical Analysis}

In this section, we analyze various structural characteristics of
the resulting anisotropic solutions, i.e., matter density, pressure
components, anisotropic factor, energy constraints, mass function,
compactness parameter, surface redshift and adiabatic index. For
this purpose, we consider observed values of mass and predicted the
radii from condition $p_r=0$ at $r=R$ for star models, i.e., SAX
J1808.4-3658, Vela X-1, PSR J0348+0432, and 4U 1608-52 \cite{24}-
\cite{26}. The values of mass, predicted radii and matter values are
presented in Table \textbf{1}. Using these values, the unknown
constants $(\alpha,\beta,\gamma)$ are calculated in Table
\textbf{2}. Moreover, the field equations (2) reduce to the Einstein
field equations for $f(\mathcal{G})=0$. By following the technique
in \cite{8}, we predict the radii and values of constants for the
considered star models which are mentioned in Table \textbf{3}. In
$f(\mathcal{G})$ theory, the variation of matter variables and EoS
parameter are shown in Figures \textbf{2} and \textbf{5},
respectively whereas, Figures \textbf{3} and \textbf{6} exhibit the
same attributes for the GR model. The plot of total mass is
presented in Figure \textbf{4} which indicates that the anisotropic
stellar structures become more massive as the radius increases. The
graphical behavior shows that energy density as well as pressure
components exhibit large values as compared to GR \cite{27a}.
Moreover, the EoS parameters lie within the interval (0,1) and
attain higher values as compared to GR \cite{11}. The positive and
regular behavior of these variables ensure the singularity-free
nature of the matter components. It is observed that anisotropic
system is more dense in modified Gauss-Bonnet gravity in comparison
to GR \cite{27a}.
\begin{table}
\caption{Approximate values of physical parameters corresponding to
$f(\mathcal{G})=\mathcal{G}^2$ model for different stellar
candidates.}
\begin{center}
\begin{tabular}{|c|c|c|c|c|}
\hline{Star Models}&{Mass $(M_{\odot})$}&{Radius $(km)$}&{$\rho_c(gm/cm^3)$}& {$p_{rc}(dyne/cm^2)$}\\
\hline {Vela X-1} & $1.77$ & $9.05$ & $1.3\times10^{15}$ & $3.6\times10^{35}$\\
\hline {SAX J1808.4-3658}& $1.43$ & $6.9$ & $2.4\times10^{15}$
& $8.4\times10^{35}$\\
\hline {4U 1608-52} & 1.74& 9.01 & $1.4\times10^{15}$
& $9.7\times10^{35}$\\
\hline {PSR J0348+0432}& 2.1 & 10.06& $1.2\times10^{15}$ &
$4.2\times10^{35}$ \\
\hline
\end{tabular}
\end{center}
\end{table}
\begin{table}
\caption{Approximate values of constants corresponding to
$f(\mathcal{G})=\mathcal{G}^2$ model for different stellar
candidates.}
\begin{center}
\begin{tabular}{|c|c|c|c|c|c|}
\hline{Star Models}&{$\frac{m}{r}$}& {$Z$}&{$\alpha$}& {$\beta$}&{$\gamma$}\\
\hline {Vela X-1} & $0.29$ & $1.4$ & $0.0046$ & $136.61$ &$0.1883$\\
\hline {SAX J1808.4-3658}& $0.31$ & $1.5$ & $0.0085$& $76.9$&$0.1728$\\
\hline {4U 1608-52} & $0.30$& $1.6$& $0.0049$
& $131.8$&$0.1726$\\
\hline {PSR J0348+0432} & $0.29$& $1.51$& $0.0039$&$131.8$
& $0.1723$\\
\hline
\end{tabular}
\end{center}
\end{table}
\begin{table}
\caption{Approximate values of physical parameters and constants
when $f(\mathcal{G})=0$.}
\begin{center}
\begin{tabular}{|c|c|c|c|c|}
\hline{Star Models}&{Mass $(M_{\odot})$}&{Radius $(km)$}&{$\alpha$}& {$16\alpha\beta\gamma$}\\
\hline {Vela X-1} & $1.77$ & $8.93$ & $0.0032$ & $2.40$\\
\hline {SAX J1808.4-3658}& $1.43$ & $6.75$ & $0.0054$
& $2.45$\\
\hline {4U 1608-52} & 1.74& 8.80 & $0.0026$
& $2.65$\\
\hline {PSR J0348+0432}& 2.10 & 9.70& $0.0031$ &
$2.20$ \\
\hline
\end{tabular}
\end{center}
\end{table}
\begin{figure}\center
\epsfig{file=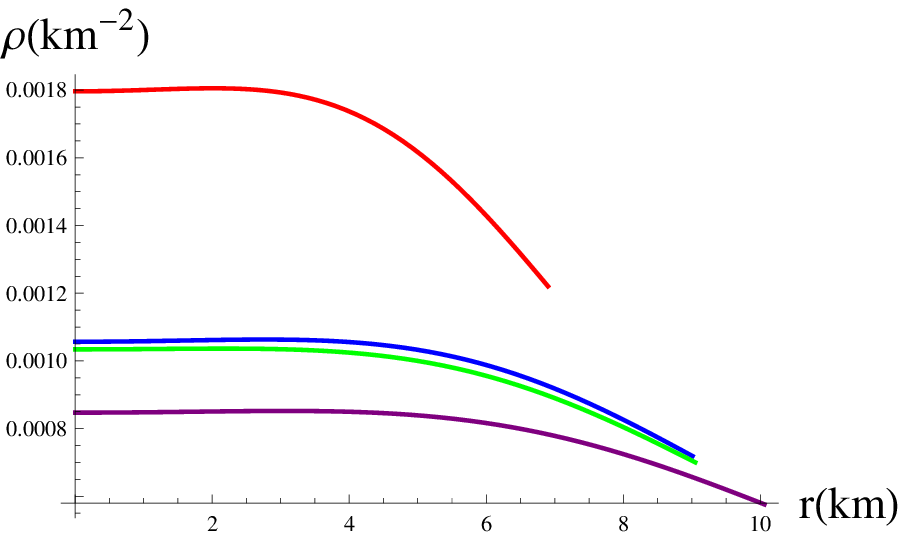,width=0.35\linewidth}
\epsfig{file=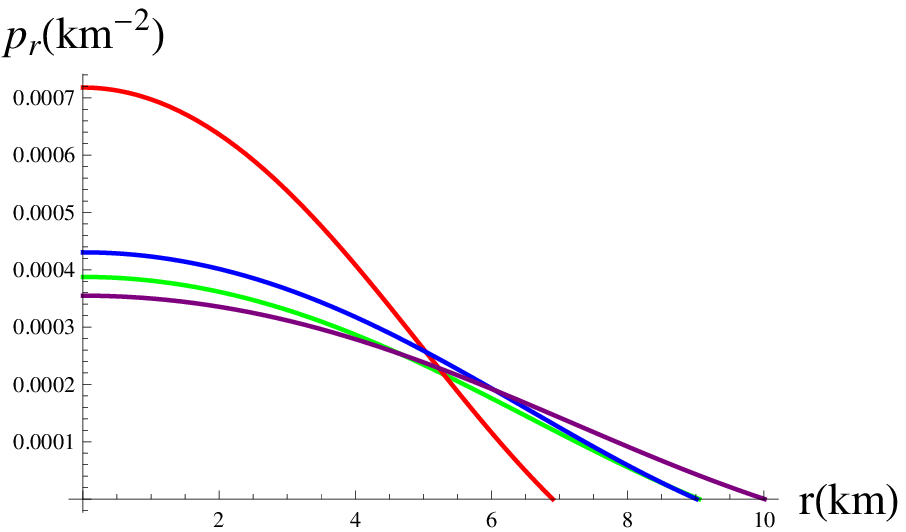,width=0.35\linewidth}
\epsfig{file=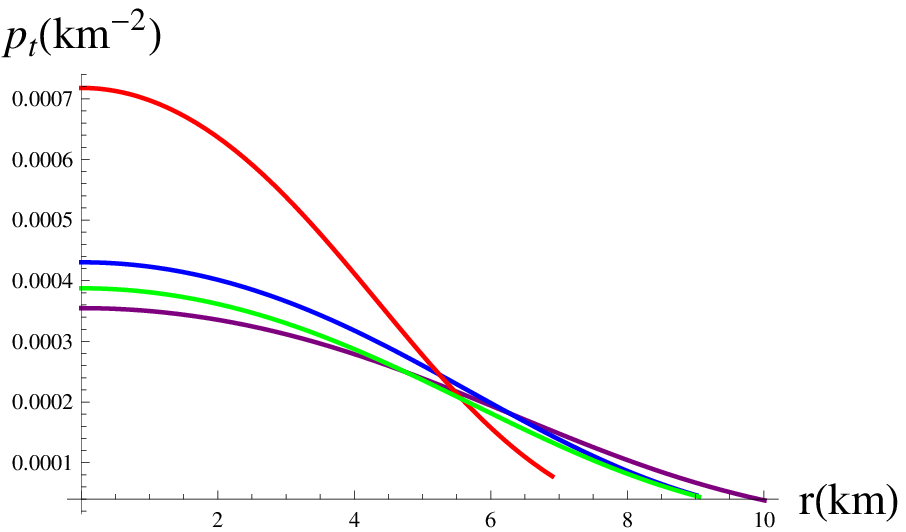,width=0.35\linewidth}
\epsfig{file=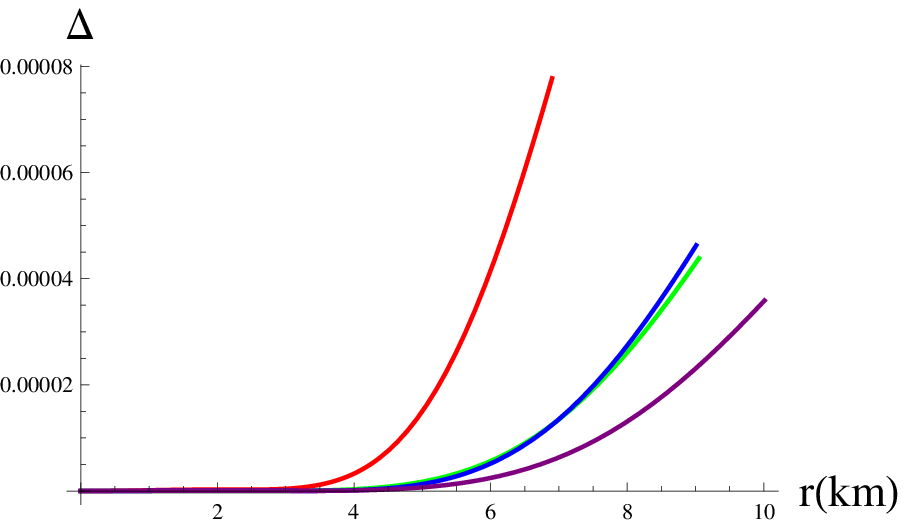,width=0.35\linewidth}\caption{Variation of
matter density, radial pressure, transversal pressure and anisotropy
for $f(\mathcal {G})=\mathcal{G}^2$ model corresponding to SAX
J1808.4-3658 (red), Vela X-1 (green), 4U 1608-52 (blue) and PSR
J0348+0432 (purple) against radial coordinate.}
\end{figure}

The pressure anisotropy helps us to analyze the matter distribution
and its direction depends on radial and transversal pressures. If
$p_{t}> p_{r}$, then anisotropy is positive which shows the outward
directed pressure whereas the case $p_{t} < p_{r}$ gives negative
anisotropy which indicates inward directed pressure. Figure
\textbf{2} shows repulsive nature of anisotropic force which is
required for stellar structures \cite{26'}. Moreover, the graphical
analysis also exhibits that anisotropy increases in the framework of
$f(\mathcal{G})$ gravity as compared to GR \cite{11}.
\begin{figure}\center
\epsfig{file=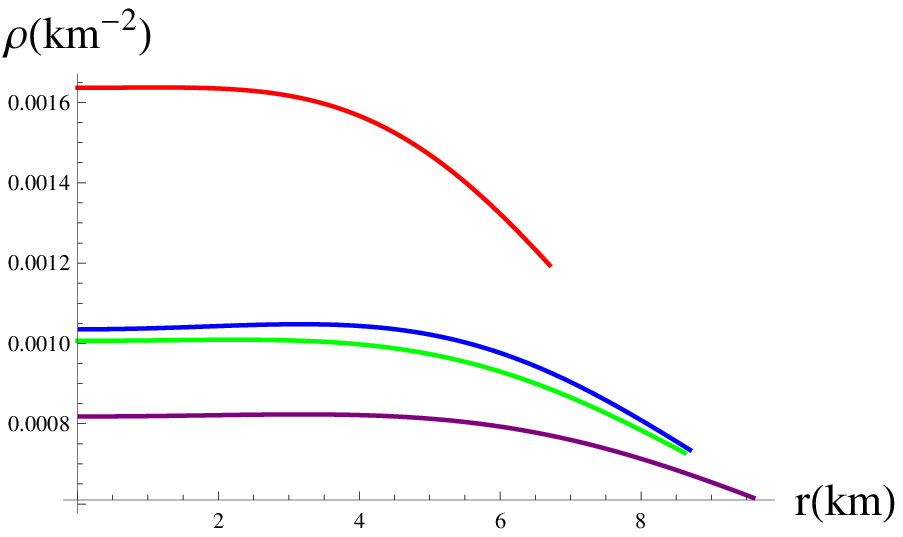,width=0.35\linewidth}
\epsfig{file=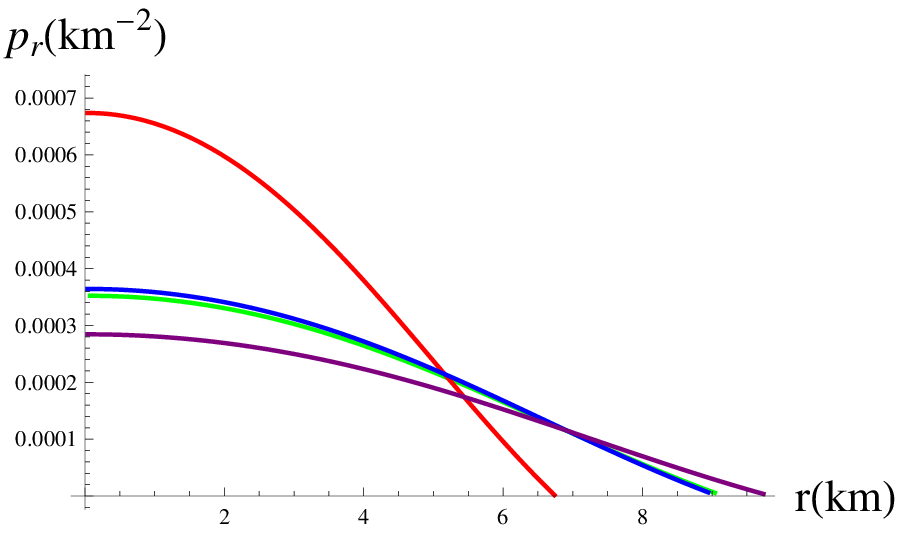,width=0.35\linewidth}
\epsfig{file=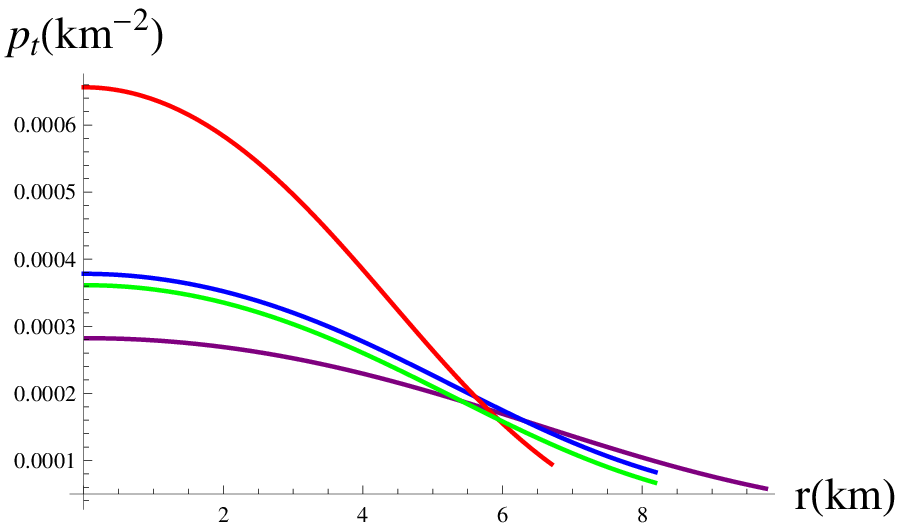,width=0.35\linewidth}
\epsfig{file=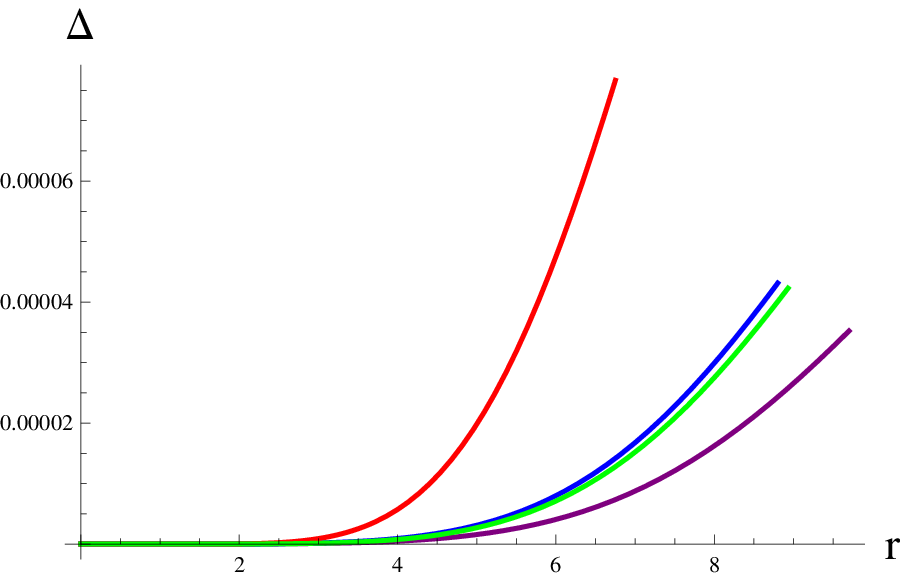,width=0.35\linewidth}\caption{Variation of
matter density, radial pressure, transversal pressure and anisotropy
for SAX J1808.4-3658 (red), Vela X-1 (green), 4U 1608-52 (blue) and
PSR J0348+0432 (purple) against radial coordinate for
$f(\mathcal{G})=0$.}
\end{figure}

\begin{figure}\center
\epsfig{file=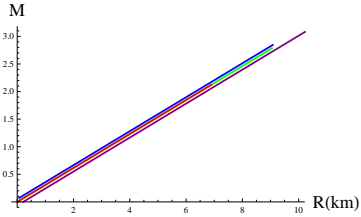,width=0.35\linewidth}\caption{Analysis of total
mass verses radius corresponding to SAX J1808.4-3658 (red), Vela X-1
(green), 4U 1608-52 (blue) and EXO 1785-248 (purple).}
\end{figure}
\begin{figure}\center
\epsfig{file=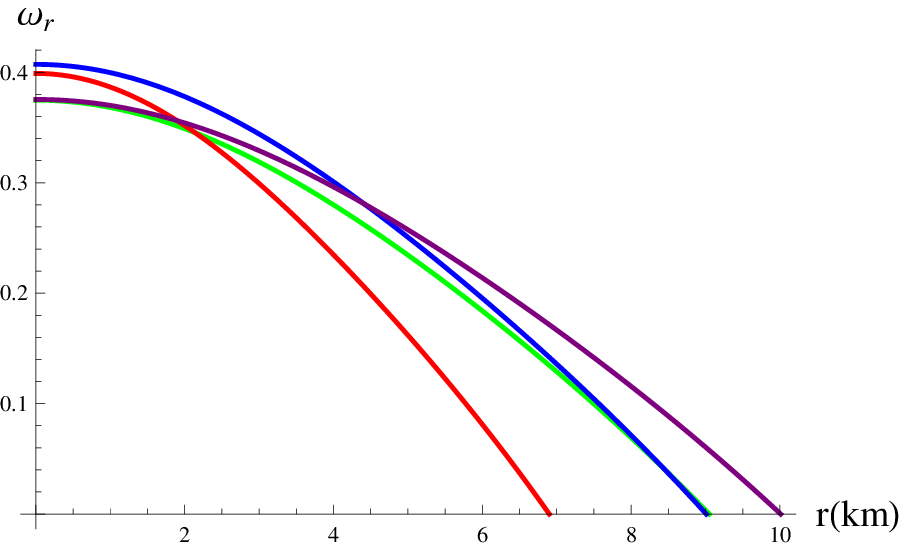,width=0.43\linewidth}
\epsfig{file=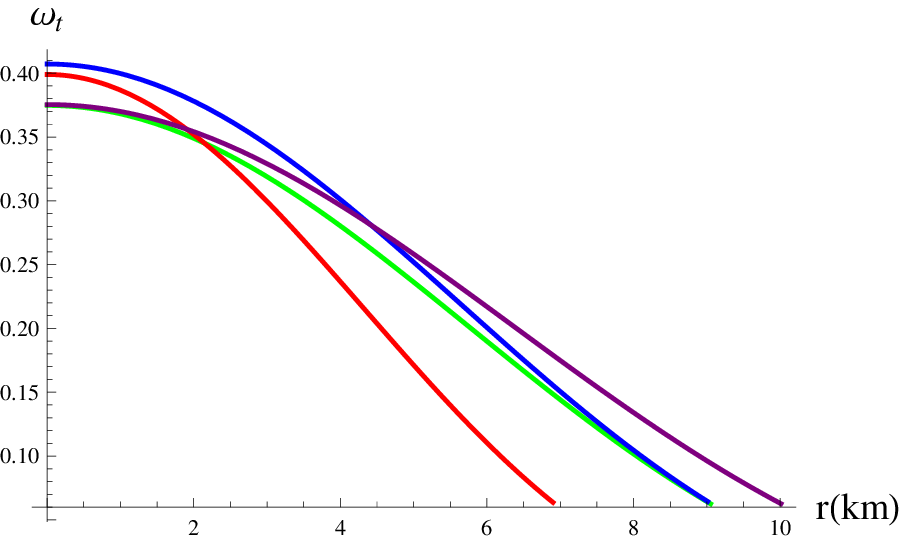,width=0.43\linewidth}\caption{Variation of EoS
parameters for $f(\mathcal {G})=\mathcal{G}^2$ model corresponding
to SAX J1808.4-3658 (red), Vela X-1 (green), 4U 1608-52 (blue) and
PSR J0348+0432 (purple) against radial coordinate.}
\end{figure}
\begin{figure}\center
\epsfig{file=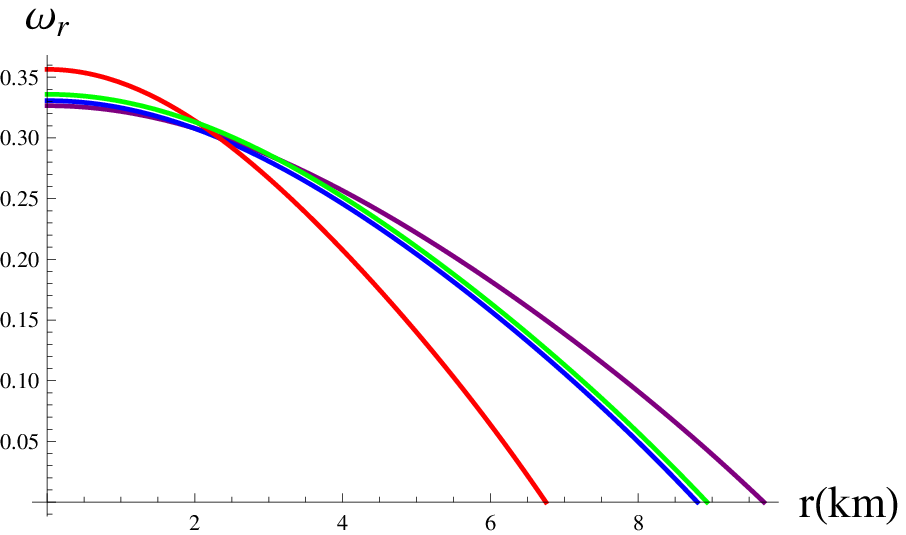,width=0.43\linewidth}
\epsfig{file=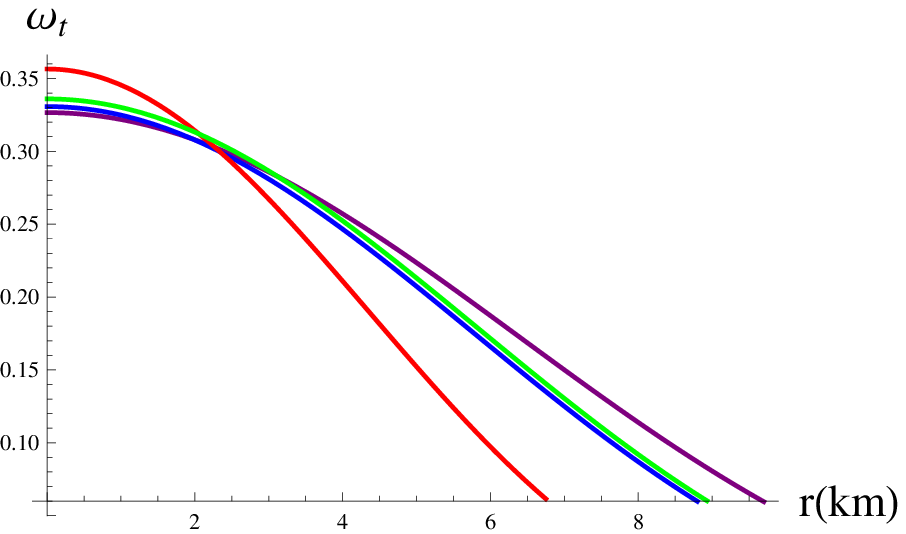,width=0.43\linewidth}\caption{Variation of EoS
parameters for SAX J1808.4-3658 (red), Vela X-1 (green), 4U 1608-52
(blue) and PSR J0348+0432 (purple) against radial coordinate when
$f(\mathcal{G})=0$.}
\end{figure}

\subsection{Energy Bounds}

The presence of ordinary or exotic matter inside the stellar
geometry is ensured by energy bounds. These conditions are
classified into null, weak, dominant and strong which must be
satisfied for the realistic existence of ordinary matter. For
anisotropic matter source, these conditions are
\begin{figure}\center
\epsfig{file=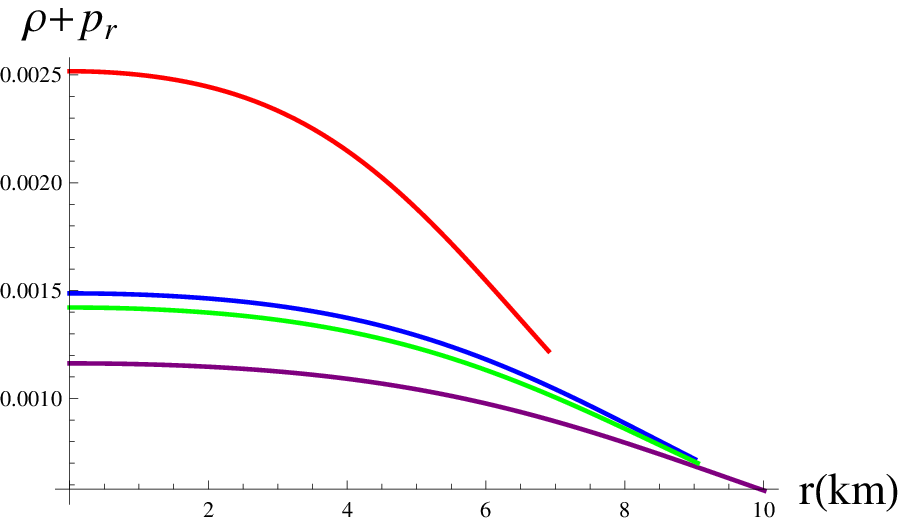,width=0.43\linewidth}
\epsfig{file=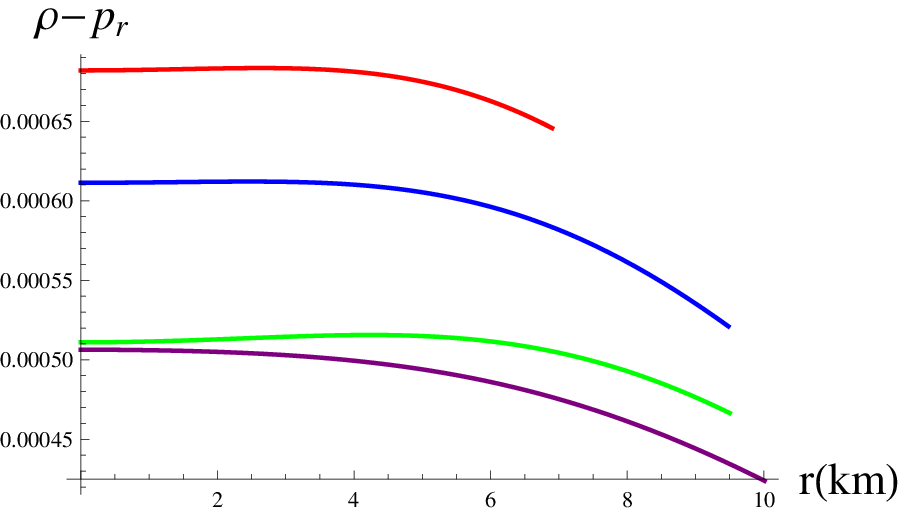,width=0.43\linewidth}
\epsfig{file=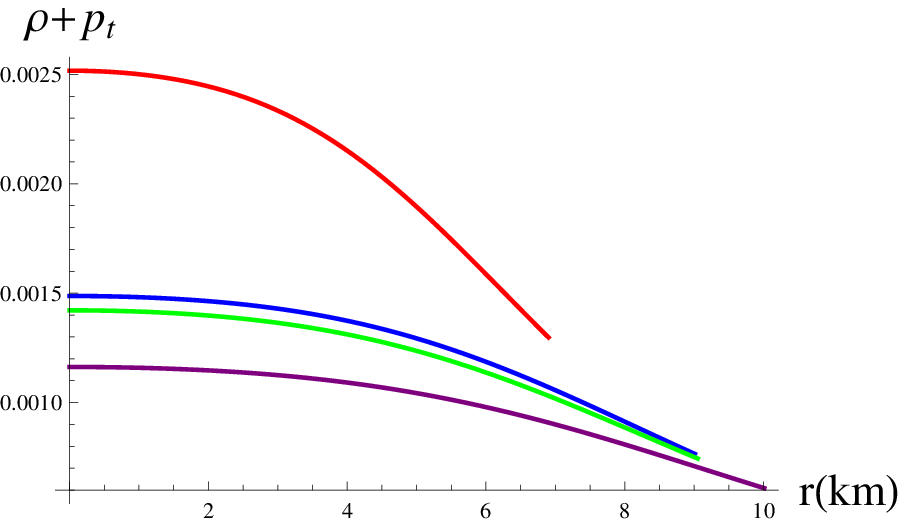,width=0.43\linewidth}
\epsfig{file=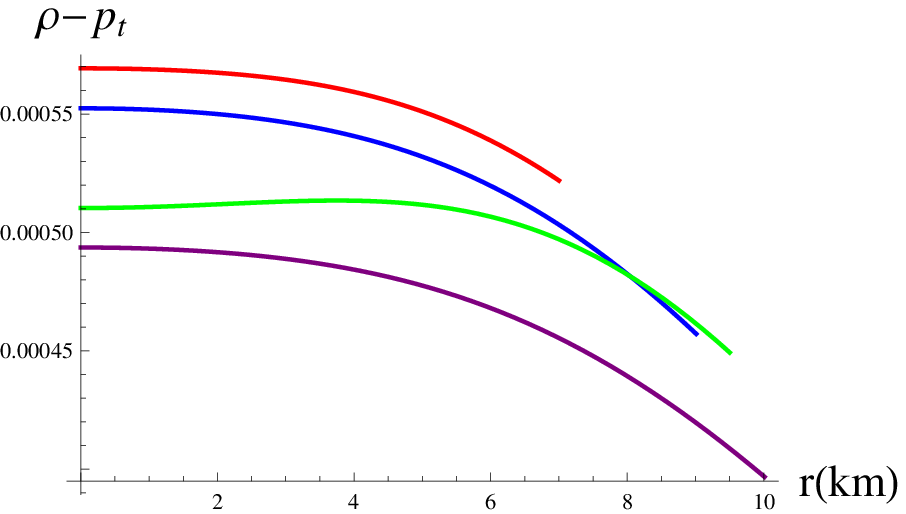,width=0.43\linewidth}
\epsfig{file=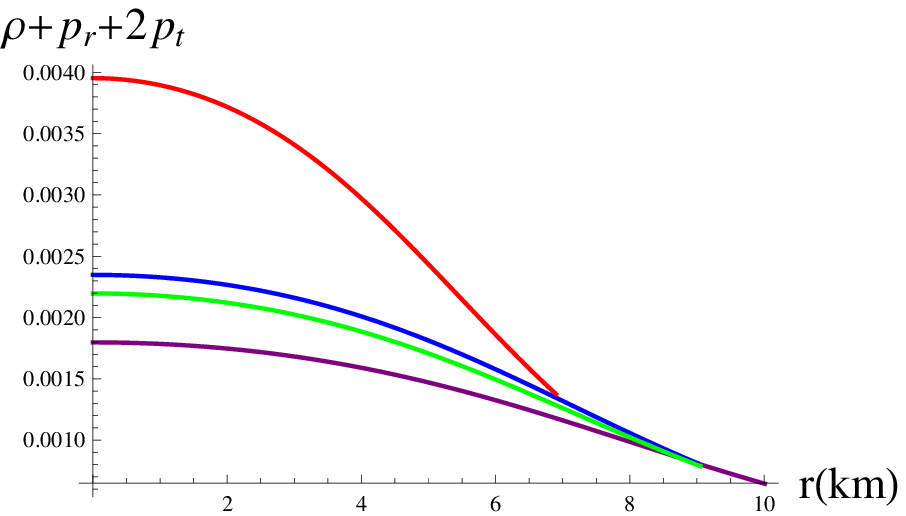,width=0.43\linewidth}\caption{Energy bounds for
$f(\mathcal {G})=\mathcal{G}^2$ model corresponding to SAX
J1808.4-3658 (red), Vela X-1 (green), 4U 1608-52 (blue) and EXO
1785-248 (purple).}
\end{figure}
\begin{figure}\center
\epsfig{file=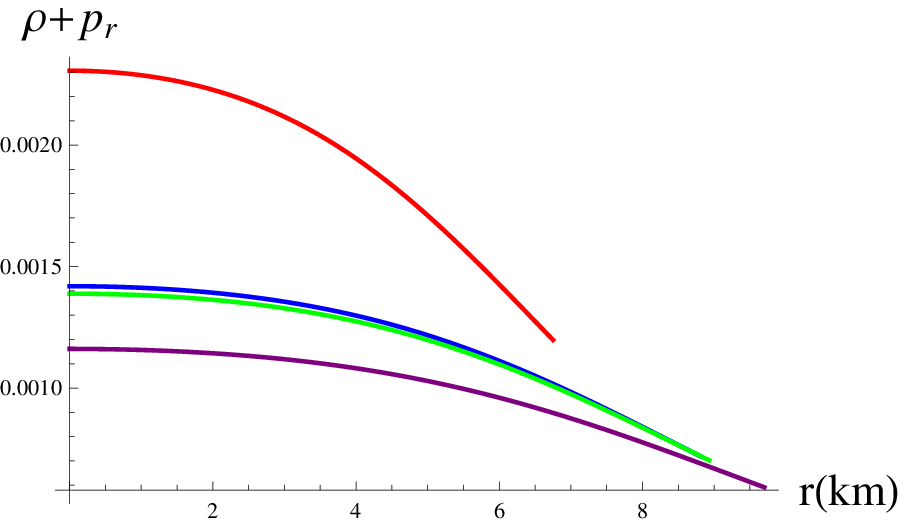,width=0.43\linewidth}
\epsfig{file=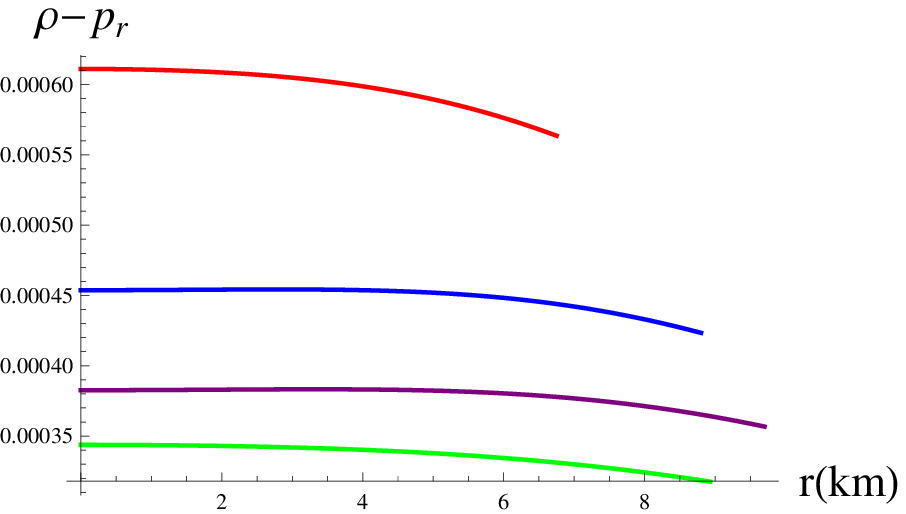,width=0.43\linewidth}
\epsfig{file=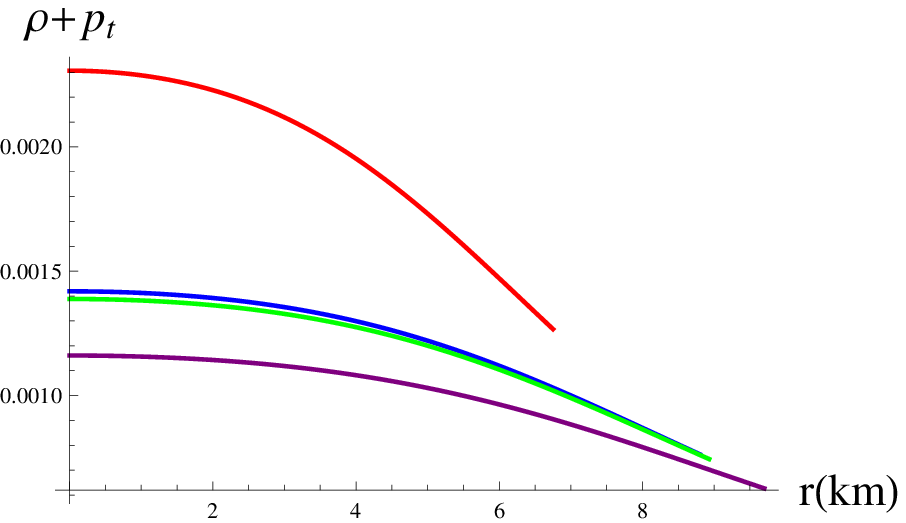,width=0.43\linewidth}
\epsfig{file=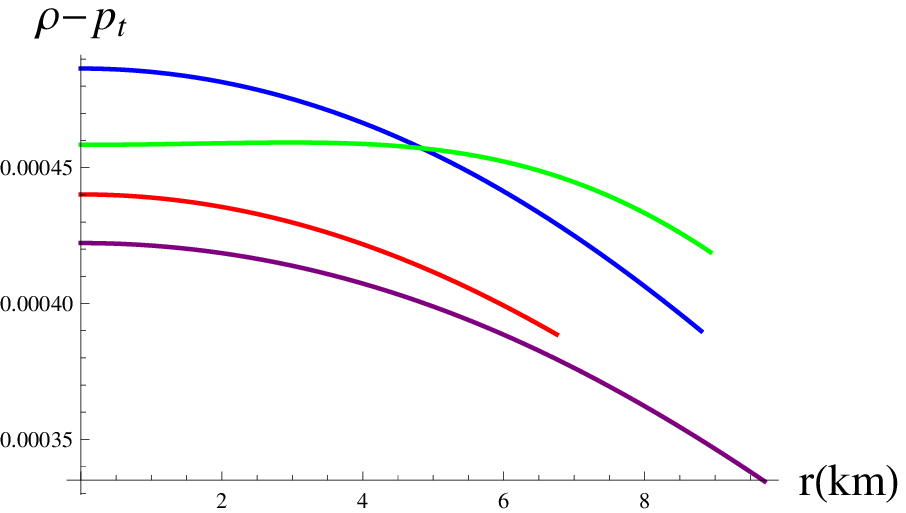,width=0.43\linewidth}
\epsfig{file=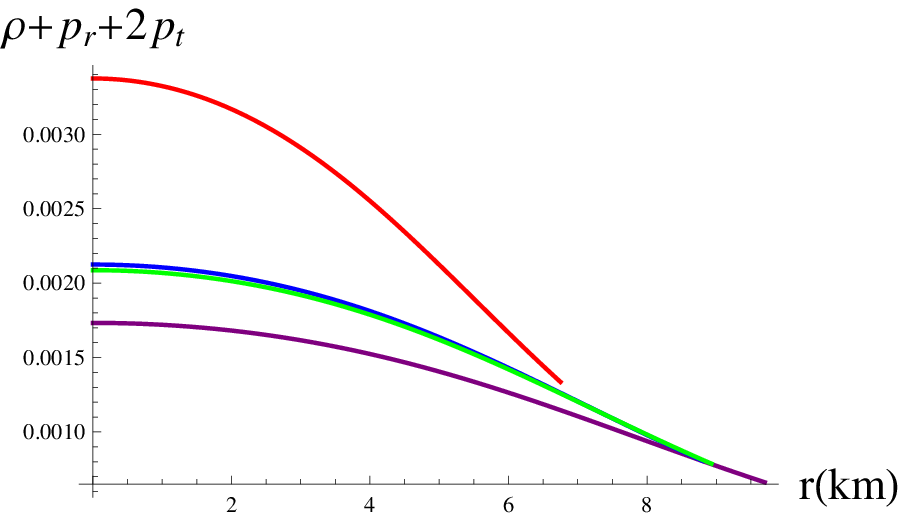,width=0.43\linewidth}\caption{Energy bounds
for SAX J1808.4-3658 (red), Vela X-1 (green), 4U 1608-52 (blue) and
EXO 1785-248 (purple) when $f(\mathcal{G})=0$.}
\end{figure}
\begin{itemize}
\item Null: \quad\quad$\rho+ p_{r}\geq 0$,
\quad $\rho+ p_{t}\geq0$,
\item Weak: \quad\quad $\rho\geq 0$,\quad$\rho
+ p_{r}\geq 0$,\quad$\rho+p_{t}\geq 0$,
\item Dominant: \quad$\rho-p_{r}\geq0$,
\quad $\rho- p_{t}\geq0$,
\item Strong:\quad$\rho+2p_{t} + p_{r}\geq 0$.
\end{itemize}
Figure \textbf{7} depicts that all the energy bounds are fulfilled
which confirm the presence of ordinary matter in compact star
models. Moreover, we also check the energy conditions when
$f(\mathcal{G})=0$. The graphical behavior (Figure \textbf{8})
confirms the presence of normal matter inside the star models. The
validity of energy conditions also leads to the physical viability
of the solution. It is worthwhile to mention here that for larger
negative values of $\chi$ energy conditions are violated in the
present work.

\subsection{Mass, Compactness and Redshift}

The mass of anisotropic stellar object through Eq.(\ref{15}) turns
out to be
\begin{equation}\label{27}
m=\frac{Mr^{3}e^{\frac{M(R^2-r^2)}{R^2(2M-R)}}}{R^2(R-2M)
+2Mr^2e^{\frac{M(R^2-r^2)}{R^2(2M-R)}}},
\end{equation}
which shows that mass and radius are inter-related quantities such
that the mass becomes zero at $r=0$. The regular behavior for mass
of the stars is verified through graphical analysis in Figure
\textbf{5}. We also observe that mass of stellar objects increases
as the radius increases. The compactness function is defined as the
mass to radius ratio for the celestial bodies given by
\begin{equation}\label{28}
\mu{(r)}=\frac{m(r)}{r}=\frac{Mr^{2}e^{\frac{M(R^2-r^2)}{R^2(2M-R)}}}
{R^2(R-2M)+2Mr^2e^{\frac{M(R^2-r^2)}{R^2(2M-R)}}}.
\end{equation}
\begin{figure}\center
\epsfig{file=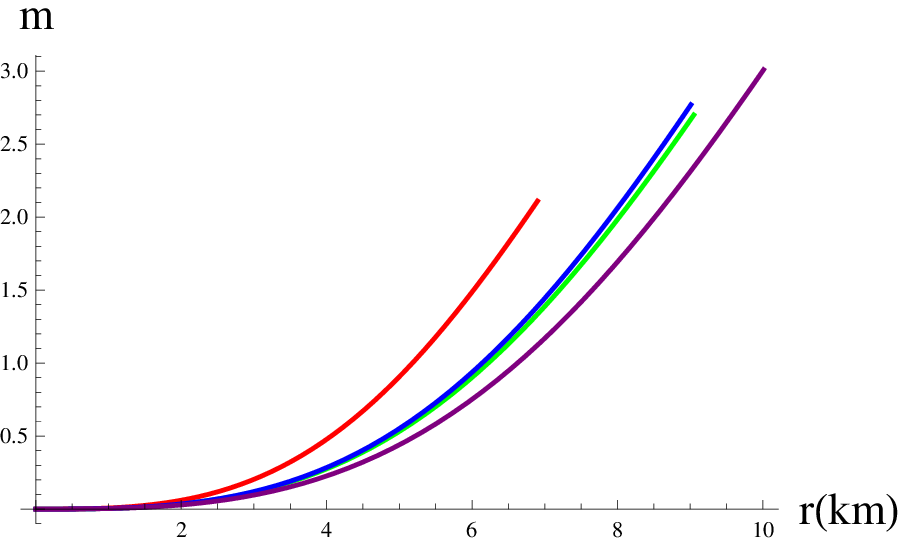,width=0.35\linewidth}
\epsfig{file=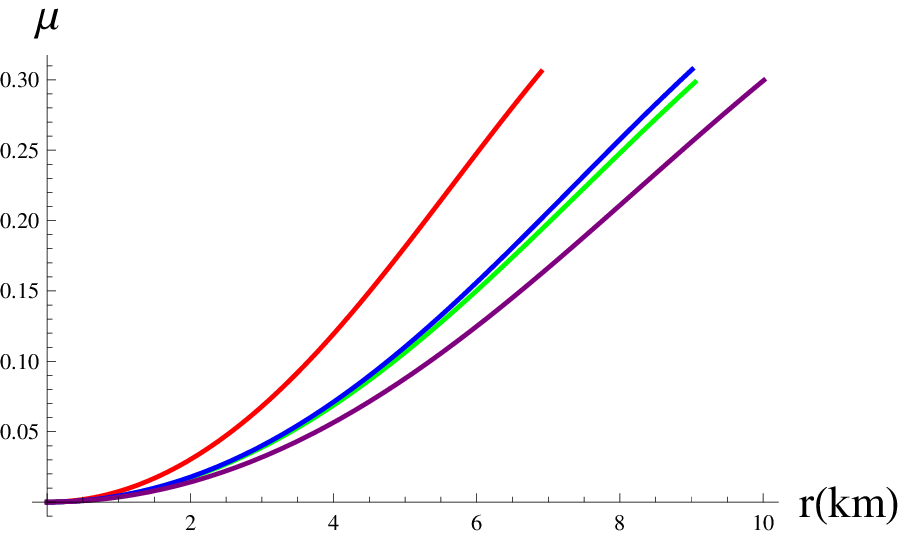,width=0.35\linewidth}
\epsfig{file=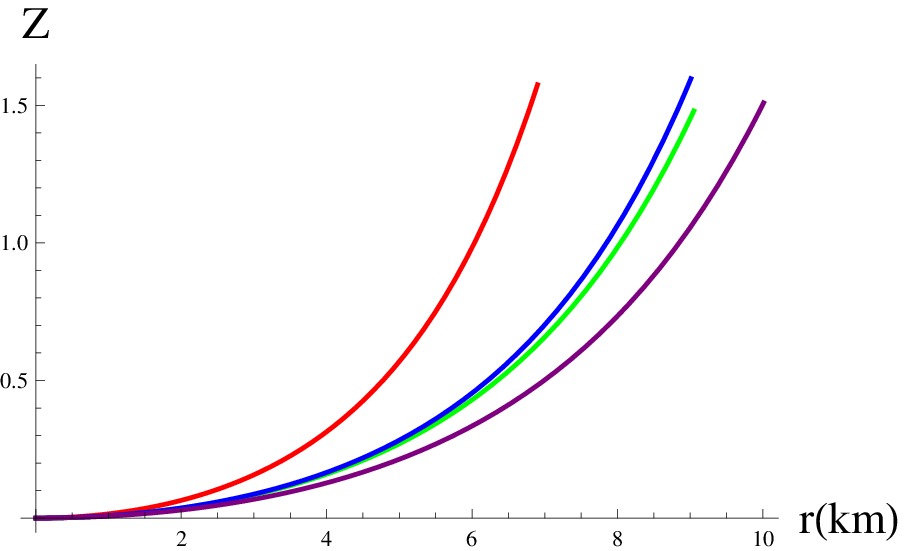,width=0.35\linewidth}\caption{Analysis of mass,
compactness parameter and gravitational redshift for $f(\mathcal
{G})=\mathcal{G}^2$ model corresponding to SAX J1808.4-3658 (red),
Vela X-1 (green), 4U 1608-52 (blue) and EXO 1785-248 (purple).}
\end{figure}
\begin{figure}\center
\epsfig{file=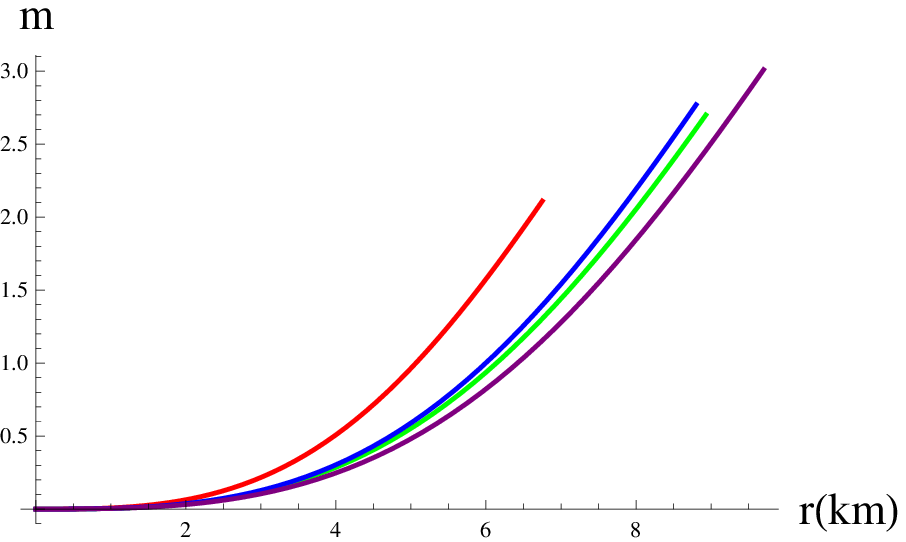,width=0.35\linewidth}
\epsfig{file=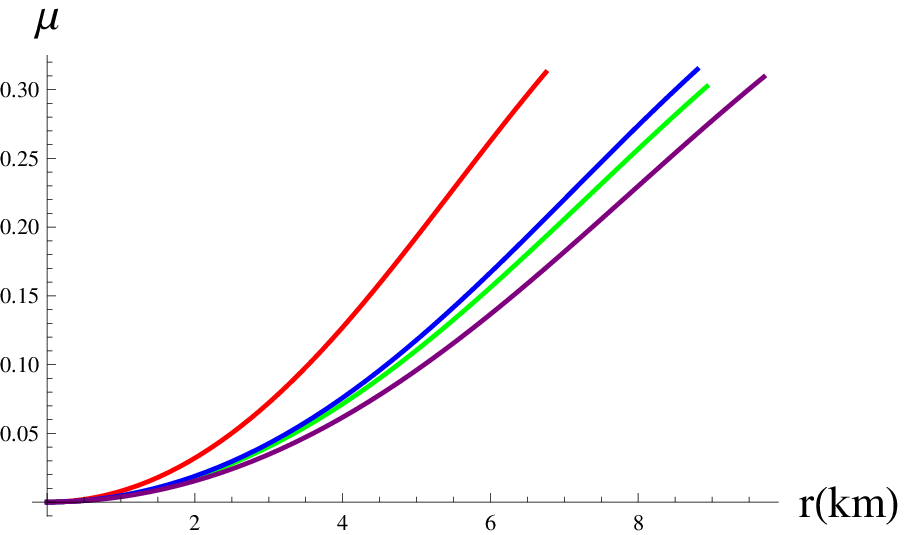,width=0.35\linewidth}
\epsfig{file=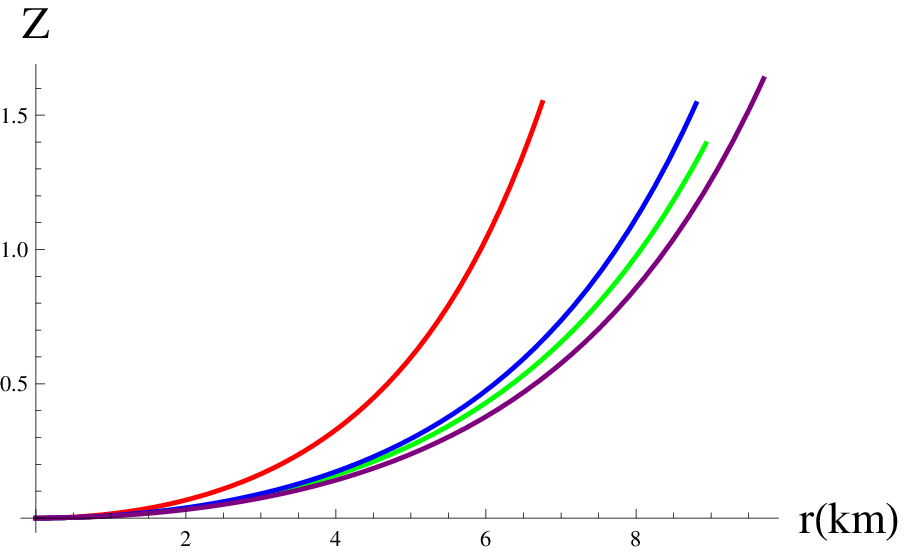,width=0.35\linewidth}\caption{Analysis of
mass, compactness parameter and gravitational redshift for SAX
J1808.4-3658 (red), Vela X-1 (green), 4U 1608-52 (blue) and EXO
1785-248 (purple) when $f(\mathcal{G})=0$.}
\end{figure}

The gravitational redshift is considered as an important element to
examine the physical behavior of celestial bodies as it measures the
force exerted on light due to strong gravity. It is defined as
$Z=-1+\frac{1}{\sqrt{1-2\mu(r)}}$, which yields
\begin{equation}\label{29}
Z=-1+\sqrt{1+\frac{2Mr^{2}e^\frac{M(R^{2}-r^{2})}{R^2(2M-R)}}{R^{2}(R-2M)}}.
\end{equation}
Figure \textbf{9} demonstrates the monotonic increasing nature of
compactness factor corresponding to different stellar models which
satisfies the Buchdahl condition ($\frac{m}{r}<\frac{4}{9}$)
\cite{27}. The gravitational redshift is also found to be
monotonically increasing function of radial coordinate and satisfies
the range for anisotropic compact stars, $Z\leq5.211$ \cite{28}.
Figure \textbf{10} exhibits that these parameters are consistent
with their respective limits for the GR model.

\subsection{Stability of Compact Stars}

Here, we discuss TOV equation, causality condition and adiabatic
index in the context of $f(\mathcal{G})$ gravity. This will help us
to examine the equilibrium and stable behavior of the obtained
solution. Astashenok et al. \cite{28a}-\cite{28d} studied the
internal structure of neutron star for different models and also
discussed the modified TOV equation in $f(\mathcal{R})$ and
$f(\mathcal{G})$ theories of gravity. The corresponding TOV equation
is constructed from the continuity equation
($\nabla^{\alpha}T_{\alpha\beta}=0$) as
\begin{equation}\label{30}
\frac{-\nu'}{2}({\rho+p_{r}})-\frac{dp_{r}}{dr}+\frac{2}{r}
({p_{t}-p_{r}})=0,
\end{equation}
where $\frac{-\nu'}{2}({\rho+p_{r}})$ depicts the gravitational
force $(F_g)$, $-\frac{dp_{r}}{dr}$ represents the hydrodynamic
force $(F_h)$ and $\frac{2}{r}({p_{t}-p_{r}})$ indicates the
anisotropic force $F_a$. Figure \textbf{11} indicates that the sum
of all these forces is zero, i.e., $F_g+F_a+F_h=0$ which assures the
existence of the equilibrium system.

A stable stellar system is considered to be more realistic from
astrophysical point of view. It is necessary to inspect the behavior
of the matter source after its departure from the state of
equilibrium, when non-vanishing radial forces of different signs
appear within the system. We examine stability of the solution
through causality condition and adiabatic index. To preserve the
causality condition, the stable anisotropic spherical objects must
have speed of sound less than that of light \cite{29}, i.e.,
$0<v^2_r<1$ and $0<v_t^2<1$, where $v_r$ and $v_t$ represent the
radial and tangential components of sound speeds, respectively
expressed as
\begin{equation}\label{31}
v_{r}^{2}=\frac{dp_{r}}{d\rho},\quad v_{t}^{2}=\frac{dp_{t}}{d\rho}.
\end{equation}
\begin{figure}\center
\epsfig{file=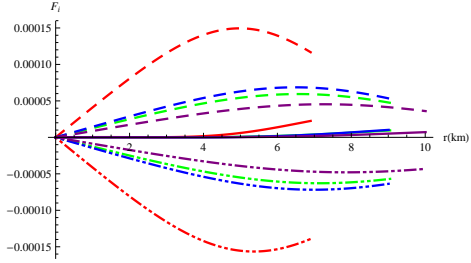,width=0.45\linewidth}\caption{Behavior of
different forces $F_{h}$ (dashed lines), $F_{a}$ (solid lines) and
$F_{g}$ (dotted dashed lines) for $f(\mathcal {G})=\mathcal{G}^2$
model corresponding to SAX J1808.4-3658 (red), Vela X-1 (green), 4U
1608-52 (blue) and PSR J0348+0432 (purple).}
\end{figure}
\begin{figure}\center
\epsfig{file=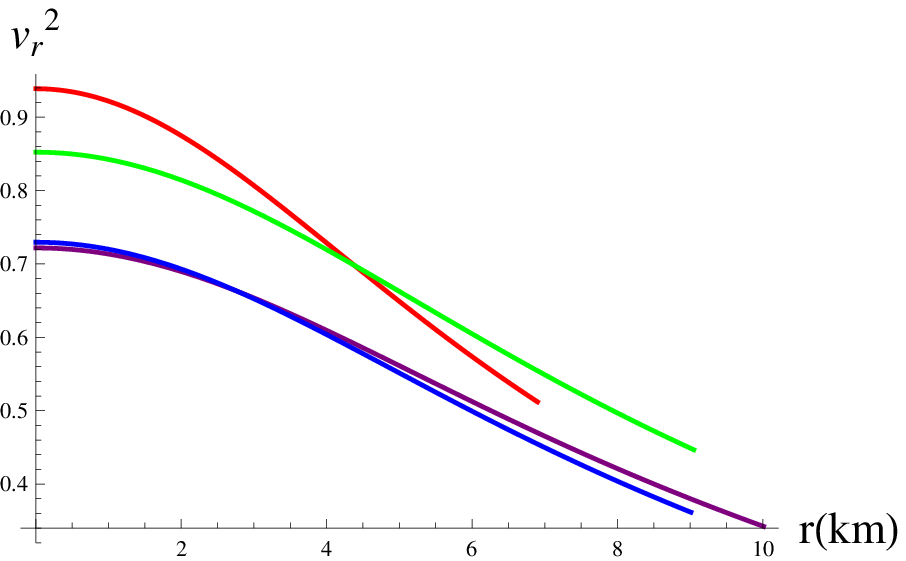,width=0.43\linewidth}
\epsfig{file=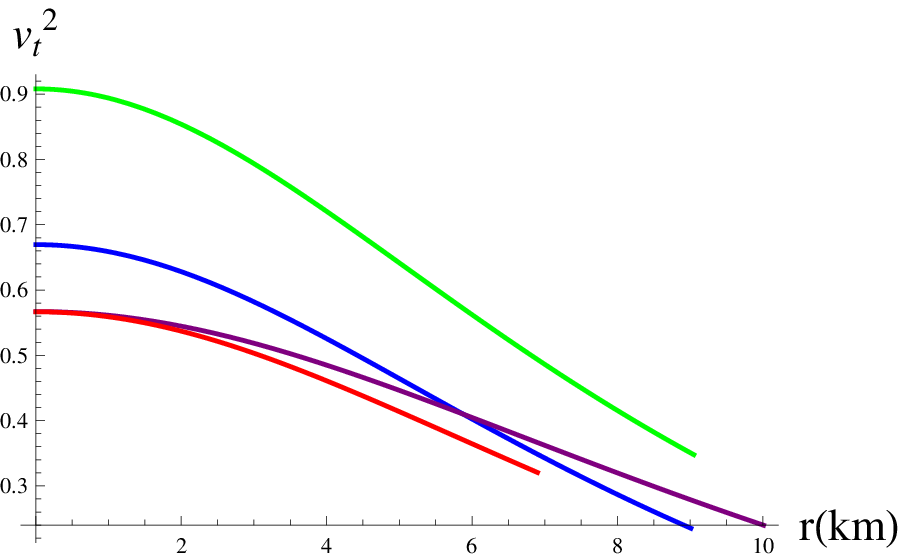,width=0.43\linewidth} \caption{Variation of
radial and tangential sound speeds for $f(\mathcal
{G})=\mathcal{G}^2$ model corresponding to SAX J1808.4-3658 (red),
Vela X-1 (green), 4U 1608-52 (blue) and PSR J0348+0432 (purple).}
\end{figure}
\begin{figure}\center
\epsfig{file=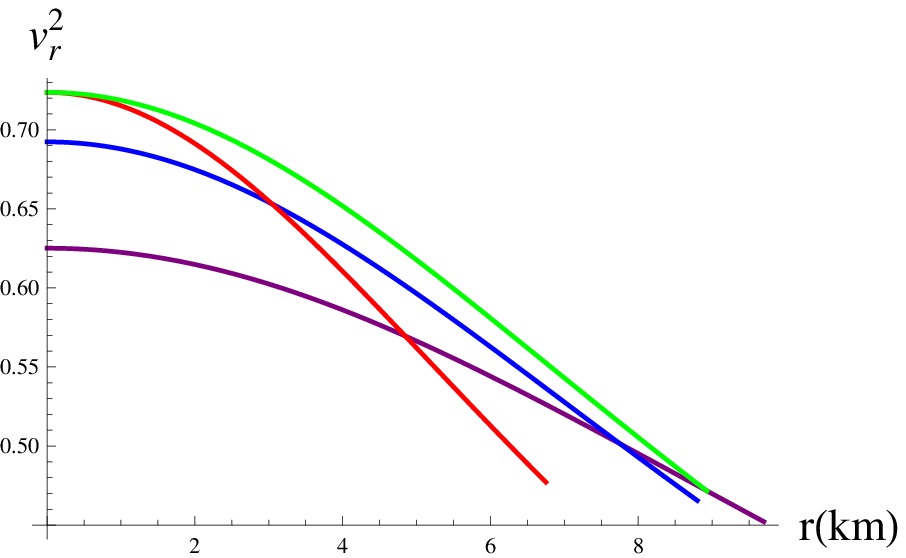,width=0.43\linewidth}
\epsfig{file=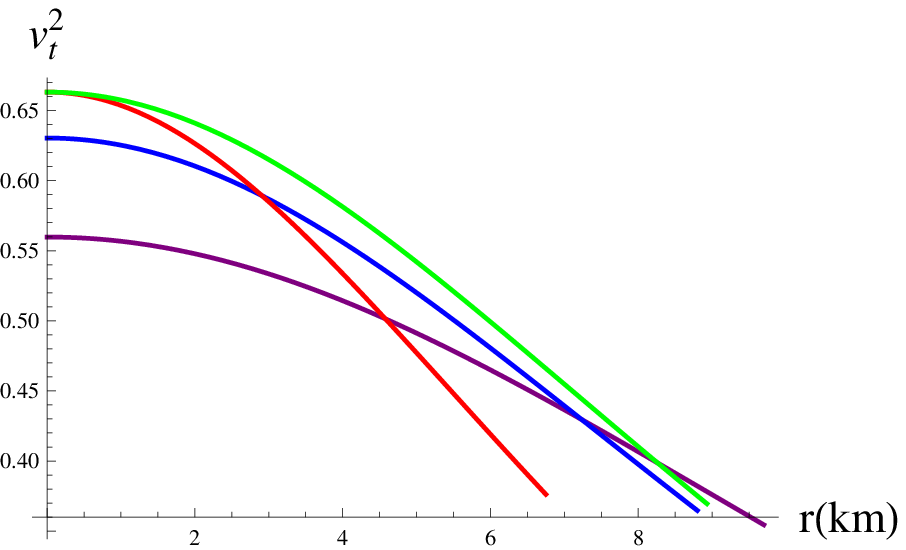,width=0.43\linewidth} \caption{Variation of
radial and tangential sound speeds for SAX J1808.4-3658 (red), Vela
X-1 (green), 4U 1608-52 (blue) and PSR J0348+0432 (purple) when
$f(\mathcal{G})=0$.}
\end{figure}

To examine the potentially stable or unstable structures of stellar
objects, we consider Herrera's cracking approach \cite{30}.
Accordingly, celestial objects should satisfy the inequality $0<\mid
v_t^2-v_r^2\mid <1$ for potentially stable model. Figures
\textbf{12} and \textbf{13} indicate that square of the radial and
tangential sound speeds satisfy the causality condition in
$f(\mathcal{G})$ theory and GR, respectively. Moreover, the plots of
absolute value of the difference of radial and tangential velocities
(Figures \textbf{14} and \textbf{15} - left panels) lie in the
prescribed bounds of Herrera's cracking approach.
\begin{figure}\center
\epsfig{file=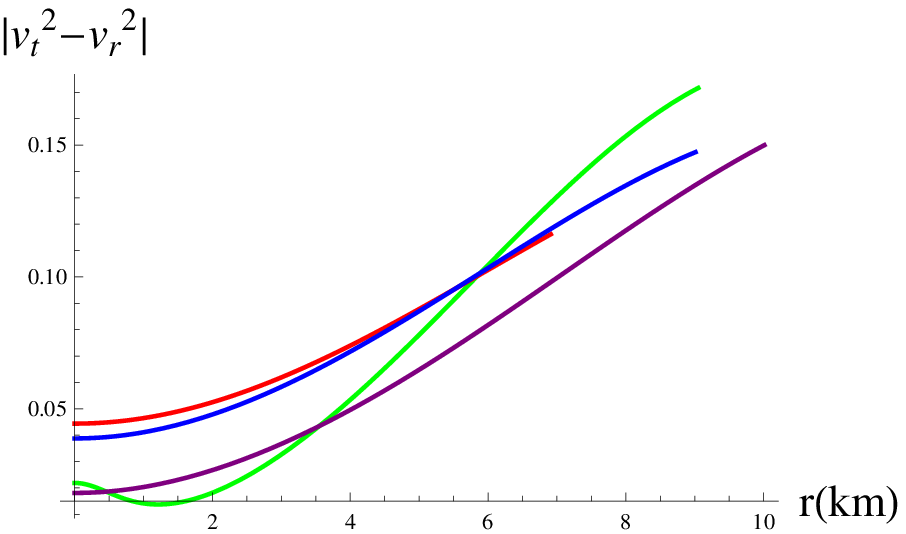,width=0.43\linewidth}
\epsfig{file=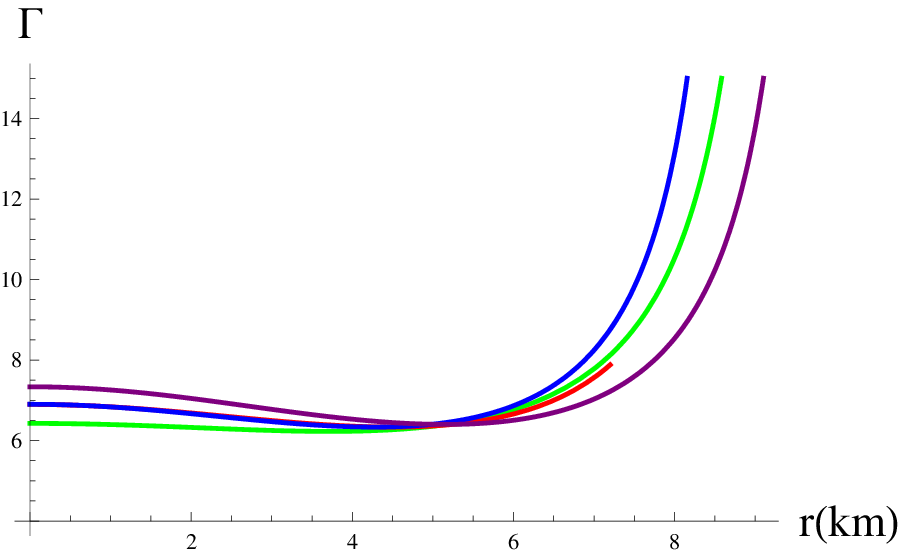,width=0.43\linewidth}\caption{Behavior of
$0<\mid v_t^2-v_r^2\mid<1$ and adiabatic index for $f(\mathcal
{G})=\mathcal{G}^2$ model corresponding to SAX J1808.4-3658 (red),
Vela X-1 (green), 4U 1608-52 (blue) and PSR J0348+0432 (purple).}
\end{figure}
\begin{figure}\center
\epsfig{file=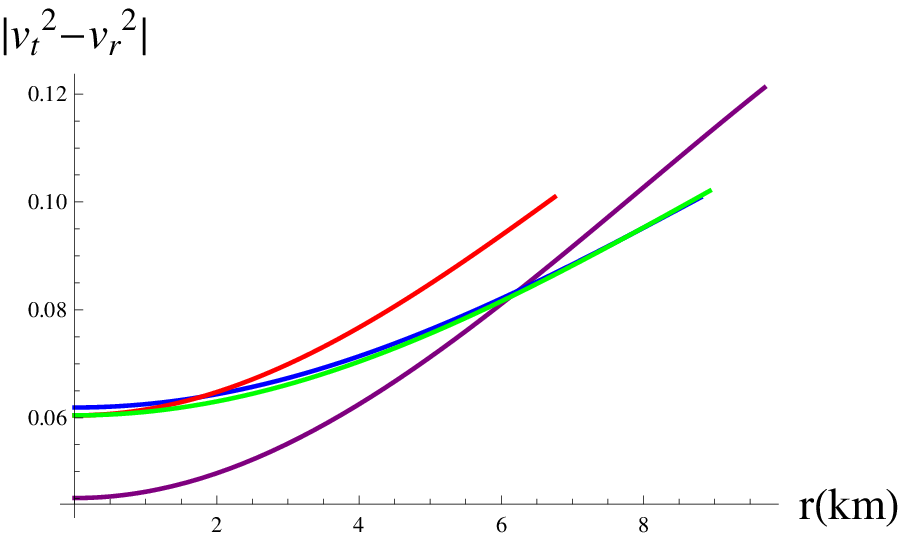,width=0.43\linewidth}
\epsfig{file=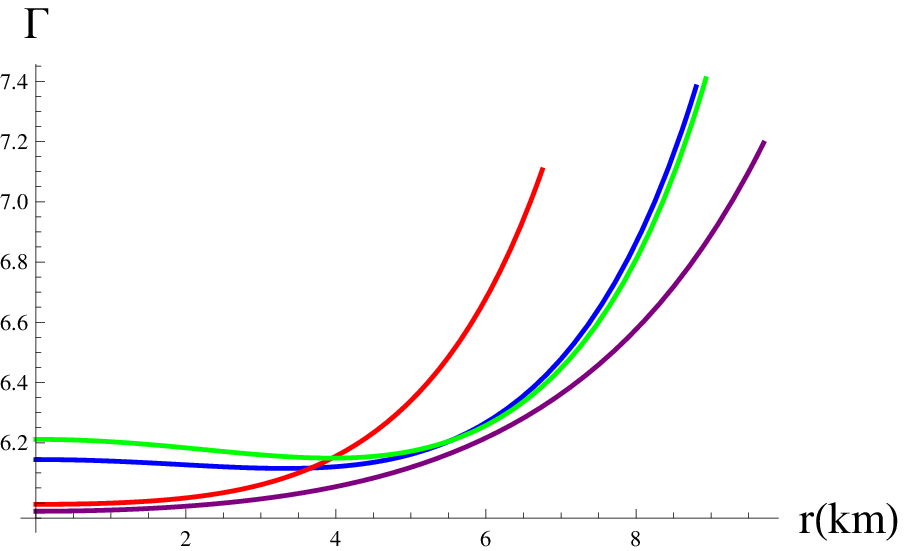,width=0.43\linewidth}\caption{Behavior of
$0<\mid v_t^2-v_r^2\mid<1$ and adiabatic index for SAX J1808.4-3658
(red), Vela X-1 (green), 4U 1608-52 (blue) and PSR J0348+0432
(purple) when $f(\mathcal{G})=0$.}
\end{figure}

For a given energy density, the adiabatic index ($\Gamma$)
illustrates the stability of relativistic as well as
non-relativistic celestial objects. The adiabatic index is a
stiffness parameter which measures the change in pressure
corresponding to a small change in density. According to Heintzmann
and Hillebrandt \cite{31}, the adiabatic index must be greater than
$\frac{4}{3}$ for the stable stellar models. The adiabatic index in
mathematical form is defined as
\begin{equation}\label{32}
\Gamma=\frac{p_{r}+\rho}{p_{r}}\frac{dp_{r}}{d\rho}=
\frac{p_{r}+\rho}{p_{r}}v_{r}^2.
\end{equation}
Figures \textbf{14} and \textbf{15} (right panels) exhibit that the
value of adiabatic index is in the defined range, i.e., greater than
$\frac{4}{3}$, for all star models. Thus, the resulting solution
shows dynamically stable behavior in $f(\mathcal{G})$ theory as well
as in GR \cite{8}.

\section{Concluding Remarks}

The purpose of this paper is to analyze the embedding class-1
solution for anisotropic spherically symmetric compact stars in
$f(\mathcal{G})$ gravity. We have smoothly matched the interior
spacetime with the exterior Schwarzschild metric to find the
arbitrary constants $(\alpha,\beta,\gamma)$. The observed masses of
SAX J1808.4-3658, Vela X-1, PSR J0348+0432 and 4U 1608-52 have been
employed to predict the radii through the condition $p_r(R)=0$. The
results are summarized as follows.
\begin{itemize}
\item We have observed that both metric functions have the lowest
values at the core of stellar models and then monotonically increase
towards the boundary. We have found that our metric functions are
compatible and fulfill all the required constraints.
\item The energy density, tangential/radial pressure
depict maximum values at the core and minimum values at the surface
of the stars which confirms the viable physical structures of
compact objects. The behavior of anisotropy is obtained such that it
is zero at the center and becomes maximum at the boundary of the
system. Moreover, EoS parameters lie in the accepted range, i.e.,
$0<\omega_r<1$ and $0<\omega_t<1$.
\item The energy conditions are satisfied assuring the presence
of normal distribution of matter inside the compact stellar
structures.
\item The compactness and redshift parameters
satisfy the required range, i.e., $\frac{m}{r}<\frac{4}{9}$ and
$Z<5.211$, respectively.
\item The physical behavior of three forces, $F_g$, $F_h$ and $F_a$
confirms the state of equilibrium for anisotropic spherical
solution.
\item Finally, we have checked stability of the system through
causality, Herrera conditions and adiabatic index. These conditions
are satisfied demonstrating that our anisotropic compact model is
stable.
\end{itemize}

We conclude that the embedding class-1 technique in $f(\mathcal{G})$
framework is compatible as all the structural attributes of stellar
models follow the physically accepted criteria. It is worth
mentioning here that anisotropic interior solutions in this theory
represent more dense structures as compared to GR \cite{8, 11, 27a}.

\end{document}